\documentclass[journal]{IEEEtran}

\usepackage{float}  

\usepackage{caption}
\usepackage{subcaption}

 \usepackage{stfloats}

\usepackage{algorithm}  
\usepackage{algpseudocode}  
\usepackage{amsmath}  

\usepackage{graphicx}  
\usepackage{color}
\usepackage{cite}

\usepackage{graphicx}  
\usepackage{epstopdf}

\usepackage{graphicx}  

\usepackage{multirow,booktabs} 
\usepackage{makecell} 
\usepackage{diagbox}

\usepackage{stfloats}
\usepackage{graphicx}
\ifCLASSINFOpdf
\else
\fi
\usepackage{amsthm,amsmath,amssymb}

\begin{document}
	
\captionsetup[figure]{labelfont={},labelformat={default},labelsep=period,name={Fig.}}

 

\captionsetup[table]{labelformat = simple, labelsep = newline, justification=centering, textfont = sc}

%
\title{GAN Based Near-Field Channel Estimation for Extremely Large-Scale MIMO Systems}


 
\author{Ming~Ye,~\IEEEmembership{Graduate Student Member,~IEEE,}~Xiao Liang,~\IEEEmembership{Member,~IEEE,} Cunhua Pan,~\IEEEmembership{Senior Member,~IEEE,}~Yinfei Xu,~\IEEEmembership{Member,~IEEE,}~Ming Jiang,~\IEEEmembership{Member,~IEEE,} and Chunguo Li,~\IEEEmembership{Senior Member,~IEEE}


\thanks{ Manuscript received February 26, 2024; revised May 24, 2024; accepted June 15, 2024.	
	This work was supported in part by the National Natural Science Foundation of China under Grants 62271137, 61871112, 62171119, 62371119, 62201137, and 62331023, in part by Key research and development plan of Jiangsu Province under Grant BE2021013-3, and in part by the Zhi Shan Young Scholar Program of Southeast University. \textit{(Corresponding author: Ming Jiang; Xiao Liang.)}}


\thanks{Ming Ye, Xiao Liang, Cunhua Pan, Yinfei Xu, Ming Jiang, and Chunguo Li are with the National Mobile Communications Research Laboratory, Southeast University, Nanjing 210096, China. Xiao Liang and Ming Jiang are also with the Pervasive Communication Research Center, Purple Mountain Laboratories, Nanjing 211111, China (e-mail: mye@seu.edu.cn; xiaoliang@seu.edu.cn; cpan@seu.edu.cn;
	yinfeixu@seu.edu.cn; jiang$_{-}$ming@seu.edu.cn; chunguoli@seu.edu.cn).} 	
}


%
%

\markboth{}%
{Shell \MakeLowercase{\textit{et al.}}: Bare Demo of IEEEtran.cls for IEEE Journals}
%



\maketitle

\begin{abstract}
Extremely large-scale multiple-input-multiple-output
(XL-MIMO) is a promising technique to achieve
ultra-high spectral efficiency for future 6G communications.
The mixed line-of-sight (LoS) and non-line-of-sight (NLoS)
XL-MIMO near-field channel model is adopted to describe
the XL-MIMO near-field channel accurately. In this paper, a
generative adversarial network (GAN) variant based channel
estimation method is proposed for XL-MIMO 
systems. Specifically, the GAN variant is developed to
simultaneously estimate the LoS and NLoS path components
of the XL-MIMO channel. The initially estimated
channels instead of the received signals are input into the
GAN variant as the conditional input to generate the 
XL-MIMO channels more efficiently. The GAN
variant not only learns the mapping from the initially estimated
channels to the XL-MIMO channels but also learns
an adversarial loss. Moreover, we combine the adversarial
loss with a conventional loss function to ensure the correct
direction of training the generator. To further enhance the
estimation performance, we investigate the impact of the
hyper-parameter of the loss function on the performance of our
method. Simulation results show that the proposed 
method outperforms the existing channel estimation
approaches in the adopted channel model. In addition, the
proposed method surpasses the
Cram$\acute{\mathbf{e}}$r-Rao lower bound (CRLB) under low pilot overhead.

\end{abstract}

\begin{IEEEkeywords}
Near-field, XL-MIMO, channel estimation, deep learning, generative adversarial network.
\end{IEEEkeywords}

%
\IEEEpeerreviewmaketitle

\section{Introduction}
%
%
%
%
\IEEEPARstart{E}{xtremely} large-scale multiple-input-multiple-output (XL-MIMO) has been widely suggested as a promising technology for future 6G communications, since it is capable of achieving 10-fold increases in spectral efficiency \cite{b1,b2,b3}.   
The number of antennas in XL-MIMO systems for future 6G communications is much larger than that in massive MIMO systems for current 5G communications. 
The sharp increase of the number of antennas in XL-MIMO systems significantly increases the Rayleigh distance (RD) that is proportional to the antenna aperture, which enhances the near-field range by orders-of-magnitude \cite{b4,b5}. 
As the RD dramatically increases, the receiver and scatterers may no longer locate in the far-field region of XL-MIMO transmitter, which renders the far-field planar wave assumption invalid.
Therefore, the XL-MIMO near-field channel should be modelled under the spherical wavefront assumption instead of the planar wavefront assumption \cite{bb6,bb7}.

For the near-field region in XL-MIMO systems, the spherical wave assumption is decided by not only the angle of departure/arrival (AoD/AoA) but also the distance of the transmitter from the scatterer \cite{b6,b7}.   
Under the planar wave assumption, the array response vector of the channel only relates to the angle. 
There have been many efficient compressive sensing (CS) based methods to accurately estimate the far-field channel,  such as the variational Bayesian inference (VBI) \cite{bb8} and  orthogonal matching pursuit (OMP) \cite{bb9,bb10} based schemes. 
However, the existing channel estimation approaches for the far-field channel model fail to accurately estimate the near-field XL-MIMO channel \cite{b8,b9,b10}, since the existing far-field channel model mismatches the near-field channel feature.

To tackle this issue, some near-field channel estimation schemes have been proposed for extremely large-scale antenna array (ELAA) based systems. 
The authors of \cite{b11} proposed a subarray-wise channel estimation approach to estimate the near-field non-stationary channel with low complexity. 
However, the 2-dimentional  distance-angle plane in \cite{b11} was uniformly divided into multiple grids to construct the near-field transform matrix, which fails to reduce the correlation among the array steering vectors. 
To exploit the sparsity of the near-field channel, a polar-domain representation based channel estimation method was proposed in  \cite{b12} by simultaneously considering both the angle and distance information. 
However, the power of nonzero elements for the near-field channel may leak to other elements when the polar-domain sparsity is considered, which leads to a power leakage effect \cite{b13}. 
The authors of \cite{b14} proposed a model based deep learning (DL) method for XL-MIMO systems by incorporating the sparsity dictionary into one neural network, but the performance needs to be improved.

The vast majority of existing near-field channel approaches  for XL-MIMO systems usually introduces the channel sparse representation scheme and utilize the CS based method such as the OMP algorithm  to estimate the near-field channel. 
The sparse representation approach leads to high computational complexity because the 2-dimentional distance-angle plane for the near-field channel needs to be partitioned into grids \cite{b14}.  The sparse representation scheme also assumes that the signal falls precisely on these grids, thus limiting the channel estimation accuracy \cite{b15}.     
Furthermore, the near-field XL-MIMO channel model adopted in the aforementioned works \cite{b11,b12,b13,b14, b15} can not describe the line-of-sight (LoS) path component of the near-field XL-MIMO channel accurately.   
Hence, these existing near-field channel estimation approaches are not capable of accurately estimating the XL-MIMO near-field channel, especially when both the transmitter and receiver are equipped with ELAAs \cite{bb16, bb17}. 
To address this problem, the authors of \cite{bb17} developed a hybrid-field channel model including both the near-field and far-field path components,  and proposed a estimation method to respectively estimate the near-field and far-field path components.
In \cite{bb18}, the far-field and near-field path components of the hybrid-field XL-MIMO channel were estimated by using support detection and the OMP algorithm, respectively. To accurately describe the near-field XL-MIMO channel, the authors of \cite{b16} proposed a near-field XL-MIMO channel model by simultaneously considering LoS and non-line-of-sight (NLoS) path components. 
A two stage channel estimation approach was also proposed in \cite{b16} by separately estimating the LoS and NLoS path components. 
However, separately estimating the LoS and NLoS path components may lead to high computational complexity. 
Moreover, the NLoS path components were estimated in \cite{b16} by using an OMP based algorithm, which limits the channel estimation performance.

The authors of \cite{b18} have proposed generative adversarial networks (GANs) as a framework in which the generator and the discriminator are adversarially trained to learn an adversarial loss. 
Recently, GANs have achieved remarkable success in image generation such as sketch generation \cite{b19} and image-to-image translation \cite{b20}. 
The key success of GANs is to learn an adversarial loss that allows the model to generate images similar to the real images. 
Since the channel matrices can be treated as two-channel images, the problem of the channel estimation in wireless communications can be regarded as an image-to-image translation problem.  
Hence, the GAN and its variants can be exploited to solve the channel estimation problem.
The loss functions in traditional DL based approaches are not well designed, which limits the channel estimation performance.  To address this issue, the GAN and its variants such as GANs \cite{b21, bb22}, conditional GAN (CGAN) \cite{b22, b23}, and Wasserstein GAN (WGAN)\cite{b24} have been successfully used to solve the problem of far-field channel estimation. 
For instance, the authors of \cite{bb22} accurately estimated the channel with few pilots by optimizing the random input of the generator in the GAN according to the received signals. 
In \cite{b22}, the received signals were fed into the CGAN as the conditional input to estimate the cascaded channels in intelligent reflecting surface assisted MIMO systems.  
The authors of \cite{b24} integrated an improved WGAN with a CGAN to enhance the channel estimation performance of the traditional least-square (LS) estimation scheme. 
The adversarial loss introduced by the GAN architecture can be used to solve the problem of loss function design in the conventional DL based schemes, which compensates for the information loss during training neural networks \cite{b23}.
However, GAN architectures in the aforementioned works are not designed for near-field channel estimation problems in XL-MIMO communication systems.

In this paper, we investigate the channel estimation based on GANs for XL-MIMO communication systems. 
To accurately describe the LoS path component of the XL-MIMO near-field channel model, we adopt the mixed LoS/NLoS XL-MIMO near-field channel model proposed in \cite{b16}.  
We propose to use the GAN variant, i.e., Pix2pix \cite{b25}, to directly estimate both the LoS and NLoS path components of the XL-MIMO near-field channel. 
Since the number of antennas at the transmitter is usually large in XL-MIMO systems, it is important to design a near-field channel estimation method with low pilot overhead. To this end, the initially estimated channels instead of the received signals are fed into the Pix2pix as the conditional input, which can make the generator generate the XL-MIMO near-field channels more efficiently.  
Thus, the proposed method can achieve satisfactory performance under low overhead.
The initially estimated channels can be obtained by initial estimation (IE). 
Moreover, the Pix2pix is developed to estimate the near-field XL-MIMO channels by adversarially training two convolutional neural networks (CNNs), i.e., the generator and the discriminator. Therefore, the Pix2pix can not only learn the mapping from the initially estimated channels to the XL-MIMO near-field channels but also learn an adversarial loss. 
The adversarial loss helps the developed Pix2pix adapt to different noise in the mixed LoS/NLoS XL-MIMO near-field channels, which improves the channel estimation performance. 
In addition, we combine a traditional loss function with the adversarial loss to ensure the correct direction of training the generator.
We call the designed Pix2pix as IE-Pix2pix. 
The proposed IE-Pix2pix outperforms the existing two stage algorithm designed for the mixed LoS/NLoS XL-MIMO near-field channel model and surpasses the Cram$\acute{\mathbf{e}}$r-Rao lower bound (CRLB) \cite{b16} when the distance between the transmitter and the receiver is small. 
Furthermore, the IE-Pix2pix achieves satisfactory performance and has better performance than the CRLB under low overhead.
Specifically, our main contributions are summarized as follows.

1) To accurately describe the XL-MIMO near-field channel, we adopt the mixed LoS/NLoS XL-MIMO near-field channel model. Then, a channel estimation method based on a GAN variant is proposed for the mixed LoS/NLoS XL-MIMO near-field channel model. 
We are the first to exploit Pix2pix for the channel estimation problem in XL-MIMO communication systems. 
The LoS path and NLoS path components can be simultaneously estimated by using the IE-Pix2pix based channel estimation method. 
Moreover, the IE-Pix2pix outperforms the existing two stage algorithm proposed for the adopted channel model and surpasses the CRLB when the distance between the transmitter and the receiver is small.

2) To generate the XL-MIMO near-field channels more efficiently, the initially estimated channels instead of the received signals are input into the proposed IE-Pix2pix as the conditional input. 
The IE-Pix2pix not only learns the mapping from the initially estimated channels to the near-field XL-MIMO channels but also learns an adversarial loss. 
The adversarial loss introduced by the GAN architecture helps the IE-Pix2pix adapt to the complicated XL-MIMO near-field channels by training two CNNs correctly.
The IE-Pix2pix can still achieve  satisfactory performance under low overhead.

3) To ensure the correct direction of training the generator, we combine a conventional loss function with the adversarial loss, which enhances the channel estimation performance. 
Moreover, we study the impact of the hyper-parameter of the conventional loss function on the channel estimation performance of the  IE-Pix2pix. 
Adjusting the values of the hyper-parameter is capable of further adapting to the mixed LoS/NLoS environments for XL-MIMO systems, which can obtain performance improvement. 


The rest of this paper is organized as follows. Section II describes the signal model and the adopted XL-MIMO near-field channel model. 
In Section III, we describe the initial channel estimation for XL-MIMO systems, and introduce the objective functions and network architecture of the proposed IE-Pix2pix. After that, we propose a channel estimation method based on the IE-Pix2pix for XL-MIMO systems. The computational complexity analysis of the proposed IE-Pix2pix is also provided in Section III.  
Numerical results are provided in Section IV. Finally, Section V reports the conclusions.

\textit{Notations}: We use bold-face  lower-case letter $\mathbf{a}$ and upper-case letter $\mathbf{A}$ to denote a vector and a matrix, respectively.  
$ \left\| \mathbf{A} \right\|_2 $ represents the $l_2$-norm of matrix $\mathbf{A}$. 
$\left( \cdot \right) ^H$, $\left( \cdot \right) ^{-1}$, $\mathbb{E}\left[ \cdot \right]$, and $ \log \left( \cdot \right) $ denote conjugate transpose, inverse, expectation, and logarithmic operators, respectively. 
$\mathcal{C} \mathcal{N} \left( \mu ,\sigma^2 \right)$ denotes circularly symmetric complex Gaussian distribution with mean $\mu$ and variance $\sigma^2$.
$\mathcal{U}\left( -a,a \right)$ represents the uniform distribution on $\left( -a,a \right)$.
$\mathbf{I}$ denotes an identity matrix. 
$\lceil \cdot \rceil$ denotes the ceiling function.

\section{System  Model}

In this section, we first introduce the signal model of the considered near-field XL-MIMO system, and then describe the adopted near-field XL-MIMO channel model.

\subsection{Signal  Model}

We consider a near-field XL-MIMO system, in which the transmitter is equipped with $N_{\mathrm{t}}$ antennas and $N_{\mathrm{t}}^{\mathrm{R}}$ radio frequency (RF) chains and the receiver is equipped with $N_{\mathrm{r}}$ antennas and $N_{\mathrm{r}}^{\mathrm{R}}$ RD chains. 
The hybrid precoding architecture is utilized at the BS. 
Let $\mathbf{H}\in \mathbb{C} ^{N_{\mathrm{r}}\times N_{\mathrm{t}}}$ be the channel matrix between the transmitter and the receiver. 
The received signal $\mathbf{y}_p\in \mathbb{C} ^{M_{\mathrm{r}}\times 1}$ in the $p$-th time slot can be formulated as  
\begin{equation}
\mathbf{y}_p=\mathbf{WHFx}_p+\mathbf{n}_p,
\end{equation}
where $\mathbf{x}_p\in \mathbb{C} ^{M_{\mathrm{t}}\times 1}$ and $\mathbf{n}_p\sim \mathcal{C} \mathcal{N} \left( \mathbf{0},\sigma ^2\mathbf{I}_{M_{\mathrm{r}}} \right)$ are the transmitted signal and the ${M_{\mathrm{r}}\times 1}$ noise vector with $\sigma$ being the noise power after combining in the $p$-th time slot, respectively. $\mathbf{F}\in \mathbb{C} ^{N_{\mathrm{t}}\times M_{\mathrm{t}}}$ and $\mathbf{W}\in \mathbb{C} ^{M_{\mathrm{r}}\times N_{\mathrm{r}}}$ denote the hybrid precoding matrix and the combining matrix, respectively.

Let $\mathbf{q}_p\in \mathbb{C} ^{N_{\mathrm{t}}\times 1}$ denote the signals transmitted by all the antennas at the transmitter in the $p$-th time slot as $\mathbf{q}_p=\mathbf{Fx}_p$. Then, the collection of the received signals in overall $ P $ time slots can be expressed as 
\begin{equation}
\mathbf{Y}=\mathbf{WHQ}+\mathbf{N},
\end{equation} 
where $\mathbf{N}=\left[ \mathbf{n}_1,\mathbf{n}_2,\cdots ,\mathbf{n}_P \right] \in\mathbb{C}^{M_{\mathrm{r}}\times P}$, $\mathbf{Y}=\left[ \mathbf{y}_1,\mathbf{y}_2,\cdots ,\mathbf{y}_P \right] \in\mathbb{C}^{M_{\mathrm{r}}\times P}$, and $\mathbf{Q}=\left[ \mathbf{q}_1,\mathbf{q}_2,\cdots ,\mathbf{q}_P \right] \in\mathbb{C}^{N_{\mathrm{t}}\times P}$ denote the noise matrix, the received signals, and the pilot matrix, respectively.  
It is essential to design a near-field channel estimation method with low pilot overhead ($P<N_{\mathrm{t}}$), since the number of antennas at the transmitter usually becomes large in XL-MIMO communication systems.
Next, we describe the near-field XL-MIMO channel model adopted in this paper.

\begin{figure}	
	\centering
	\includegraphics[width=8cm]{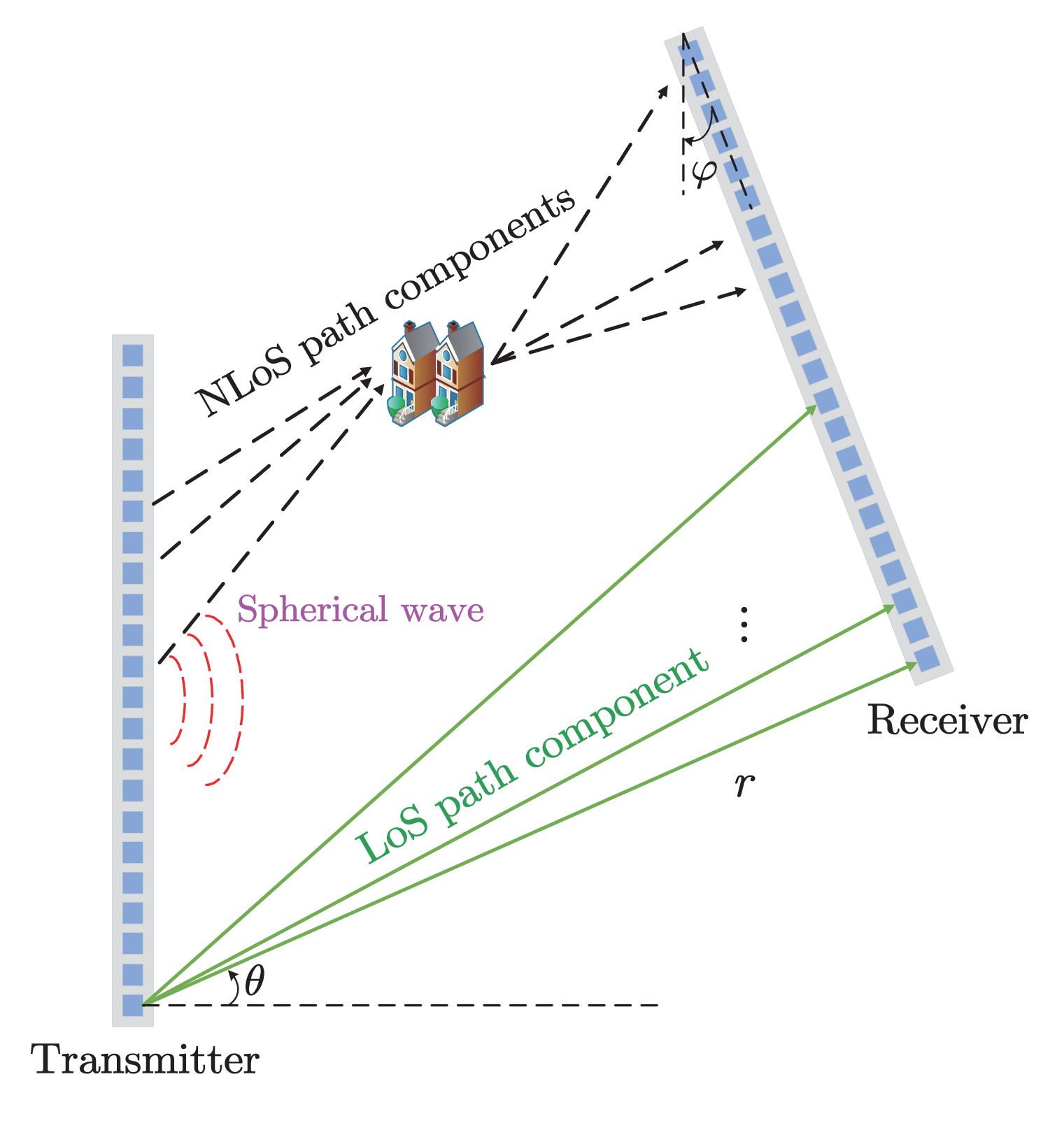}
	\caption{The adopted near-field channel model for XL-MIMO systems.}
	\label{fig.2}
\end{figure}

\subsection{The Adopted Near-Field XL-MIMO Channel Model}

Since the existing near-field channel model mismatches the practical feature of the near-field XL-MIMO LoS path component,  it is not suitable to use the near-field response vectors to model the XL-MIMO near-field channel. 
Therefore, we adopt the mixed LoS/NLoS XL-MIMO near-field channel model proposed in \cite{b16}. Different from the existing XL-MIMO near-field channel model, the XL-MIMO near-field channel model proposed in \cite{b16} contains not only the NLoS path components but also the LoS path component.

Let $d$, $\alpha_l$, and $ L$ denote the antenna spacing, the complex gain at the $ l $-th path, and the number of NLoS path components, respectively. 
Let $\theta _{\mathrm{r},l}$ and $\theta _{\mathrm{t},l}$ denote the angles of the $ l $-th path at the receiver and the transmitter, respectively. 
$d_{l}^{\mathrm{t}}$ ($d_{l}^{\mathrm{r}}$) denotes the distance between the center of the antenna array of the transmitter (receiver) and the $l$-th scatterer. 
For uniform linear arrays (ULAs), $\mathbf{b}\left( \theta _{\mathrm{r},l},d_{l}^{\mathrm{r}} \right)$ and $\mathbf{b}\left(\theta _{\mathrm{t},l},d_{l}^{\mathrm{t}} \right)$ denote the near-field array response vectors at the receiver and the transmitter under the spherical wave assumption, respectively. 
Let $d_{l}^{\mathrm{t}}\left( u \right) , u=1,2,\cdots ,N_{\mathrm{t}},$ be the distance between the $ u $-th element on the transmitter antenna array and the $ l $-th scatterer as $ d_{l}^{\mathrm{t}}\left( u \right) =\sqrt{\left( d_{l}^{\mathrm{t}} \right)^2+d^2\delta _{u}^{2}-2dd_{l}^{\mathrm{t}}\delta _u\sin \theta _{\mathrm{t},l}}$, where $ \delta _u=\small{\frac{2u-N_{\mathrm{t}}-1}{2}}$. 
Let $d_{l}^{\mathrm{r}}\left( v \right) , v=1,2,\cdots ,N_{\mathrm{r}},$ denote the distance between the $v$-th element on the receiver antenna array and the $l$-th scatterer as $ d_{l}^{\mathrm{r}}\left( v \right) =\sqrt{\left( d_{l}^{\mathrm{r}} \right)^2+d^2\delta _{v}^{2}-2dd_{l}^{\mathrm{r}}\delta_v\sin \theta _{\mathrm{r},l}}$, where $ \delta_v=\small{\frac{2v-N_{\mathrm{r}}-1}{2}}$.   
Then, the XL-MIMO near-field channel \cite{b16} only containing NLoS path components can be expressed as 
\begin{equation}
\mathbf{H}_{\mathrm{NLoS}}=\sum_{l=1}^L{\alpha _l}\mathbf{b}\left( \theta _{\mathrm{r},l},d_{l}^{\mathrm{r}} \right) \mathbf{b}^H\left( \theta _{\mathrm{t},l},d_{l}^{\mathrm{t}} \right),
\end{equation}
where 
\begin{equation}
\mathbf{b}\left( \theta _{\mathrm{r},l},d_{l}^{\mathrm{r}} \right) =\small{\frac{1}{\sqrt{N_{\mathrm{r}}}}}\left[ e^{-j\small{\frac{2\pi}{\lambda}\left( d_{l}^{\mathrm{r}}\left( 1 \right) -d_{l}^{\mathrm{r}} \right)}},\cdots ,e^{-j\small{\frac{2\pi}{\lambda}\left( d_{l}^{\mathrm{r}}\left( N_{\mathrm{r}} \right) -d_{l}^{\mathrm{r}} \right)}} \right] ^H,
\end{equation}
and
\begin{equation}
\mathbf{b}\left( \theta _{\mathrm{t},l},d_{l}^{\mathrm{t}} \right) =\small{\frac{1}{\sqrt{N_{\mathrm{t}}}}}\left[ e^{-j\small{\frac{2\pi}{\lambda}\left( d_{l}^{\mathrm{t}}\left( 1 \right) -d_{l}^{\mathrm{t}} \right)}},\cdots ,e^{-j\small{\frac{2\pi}{\lambda}\left( d_{l}^{\mathrm{t}}\left( N_{\mathrm{t}} \right) -d_{l}^{\mathrm{r}} \right)}} \right] ^H.
\end{equation}

Different from the LoS path component, the NLoS path components of the XL-MIMO near-field channel can be changed into the polar-domain channel by utilizing the polar-domain transform matrix \cite{b12}. 
The free space propagation assumption \cite{b26} is employed to accurately model the LoS path component of the XL-MIMO near-field channel for every transmitter-receiver pair.
The LoS path component of the XL-MIMO near-field channel between the $ u $-th antenna at the transmitter and the $ v $-th antenna at the receiver can be expressed as   
\begin{equation}
\mathbf{H}\left( v,u \right) =\frac{1}{r_{v,u}}e^{-j\small{\frac{2\pi}{\lambda}r_{v,u}}},
\end{equation}
where $r_{v,u}$ represents the distance between the $ u $-th antenna at the transmitter and the $ v $-th antenna at the receiver.

As shown in Fig. 1, $r$ denotes the distance between the 1-st antenna at the transmitter and the 1-st antenna at the receiver. 
In Fig. 1, blue squares at the transmitter and the receiver correspond to the antenna array elements at the transmitter and the receiver, respectively. 
Let $\theta$ denote the AoD of the signal from the transmitter to the receiver and $\varphi$  denote the relative angle between the transmitter and the receiver. 
According to \cite{b16, b26}, the $r_{v,u}$ is given by 
\begin{equation}
\begin{aligned}
r_{v,u}&=\sqrt{\left( r\cos \theta -d_2\sin \varphi \right) ^2+\left( r\sin \theta +d_2\cos \varphi -d_1 \right) ^2}\\
&=\sqrt{\varDelta +2\left( rd_2\sin \left( \theta -\varphi \right) -rd_1\sin \theta -d_1d_2\cos \varphi \right)},\\
\end{aligned}
\end{equation}
where $\varDelta =r^2+d_{1}^{2}+d_{2}^{2}$, $d_1=ud$, and $d_2=vd$. Note that the path loss of every transmitter-receiver pair is modelled as $\frac{1}{r_{v,u}}$.

According to the geometry relation in free space, the LoS path component of the XL-MIMO near-field channel can be formulated as 
\begin{equation}
\mathbf{H}_{\mathrm{LoS}}=\mathbf{H}_{\mathrm{LoS}}\left( r,\theta ,\varphi \right) =\left[ \small{\frac{1}{r_{v,u}}e^{-j\frac{2\pi}{\lambda}r_{v,u}}} \right] _{N_{\mathrm{r}}\times N_{\mathrm{t}}}.
\end{equation}

Finally, the XL-MIMO near-field channel consisting of both the LoS and NLoS path components can be expressed as 
\begin{equation}
\begin{aligned}
\mathbf{H}=&\mathbf{H}_{\mathrm{LoS}}+\mathbf{H}_{\mathrm{NLoS}}\\
=&\mathbf{H}_{\mathrm{LoS}}\left( r,\theta ,\varphi \right)+\sum_{l=1}^L{\alpha _l}\mathbf{b}\left( \theta _{\mathrm{r},l},d_{l}^{\mathrm{r}} \right) \mathbf{b}^H\left( \theta _{\mathrm{t},l},d_{l}^{\mathrm{t}} \right).\\
\end{aligned}
\end{equation}

According to \cite{b16} and \cite{b12}, the advanced RD (ARD) and RD are defined as $\small{\frac{4D_{\mathrm{t}}D_{\mathrm{r}}}{\lambda}}$ and $\small{\frac{2\left( D_{\mathrm{t}}+D_{\mathrm{r}} \right) ^2}{\lambda}}$, where $D_{\mathrm{t}}=\small{\frac{\lambda N_{\mathrm{t}}}{2}}$ and $D_{\mathrm{r}}=\small{\frac{\lambda N_{\mathrm{r}}}{2}}$.
$D_{\mathrm{t}}$ and $D_{\mathrm{r}}$ denote the apertures of the antenna array at the transmitter and the receiver, respectively. 
The boundary between the existing XL-MIMO near-field channel model and the mixed LoS/NLoS XL-MIMO near-field channel model adopted in this paper is provided in Fig. 2. 
When the distance $r$ is smaller than the ARD, the receiver is in the mixed LoS/NLoS near-field region of the transmitter, while the receiver is in the NLoS near-field region with the distance $r$ increasing to a certain extent. 
We also show the boundary between the existing XL-MIMO near-field channel model and the far-field MIMO channel model in Fig. 2.
The receiver is in the far-field region when the distance $r$ is bigger than the RD. 
In this paper, we mainly focus on the channel estimation in the mixed LoS/NLoS near-field region.

\begin{figure*}
	\centering
	\includegraphics[width=14cm]{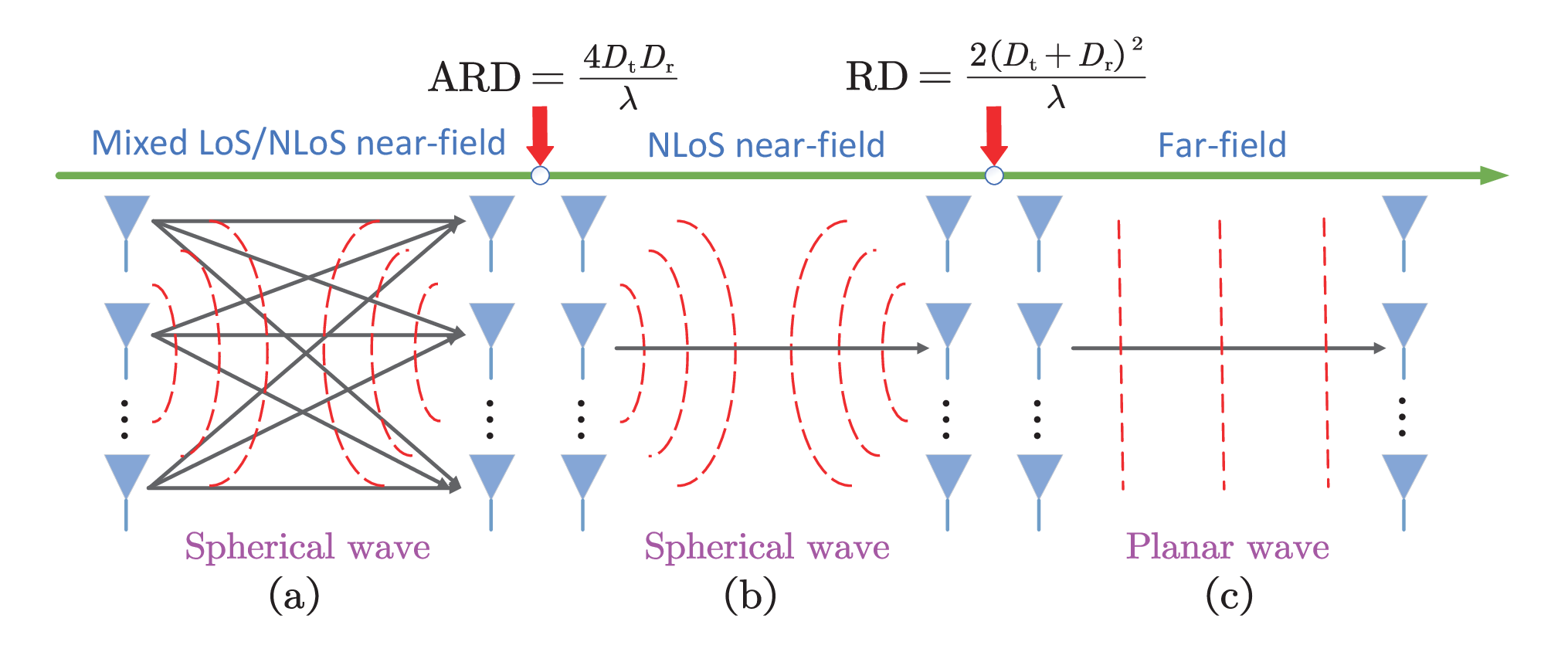}
	\caption{The near-field region and the far-field region \cite{b16}.}
	\label{fig.3}
\end{figure*}

\section{IE-Pix2pix Based Channel Estimation}

In this section, we first describe the initial channel estimation. 
Then, the objective functions and network architecture of the proposed IE-Pix2pix based on the initial channel estimation are provided. After that, we propose a channel estimation method based on the IE-Pix2pix for XL-MIMO systems. 
Finally, the computational complexity of the proposed IE-Pix2pix based method is analyzed.

\subsection{Initial Channel Estimation}

In order to estimate the near-field XL-MIMO channels more efficiently, we perform the initial channel estimation by exploiting two matrices to process the received signals in (2), yielding 
\begin{equation}
\mathbf{H}_{\mathrm{IE}}=\mathbf{G}_{\mathrm{L}}\mathbf{YG}_{\mathrm{R}}=\mathbf{G}_{\mathrm{L}}\mathbf{WHQG}_{\mathrm{R}}+\mathbf{G}_{\mathrm{L}}\mathbf{NG}_{\mathrm{R}},
\end{equation}
where $\mathbf{G}_{\mathrm{L}}=\mathbf{W}^H\left( \mathbf{WW}^H \right) ^{-1}$ and $\mathbf{G}_{\mathrm{R}}=\mathbf{Q}^H\left( \mathbf{QQ}^H \right) ^{-1}$.
After the IE, the initially estimated channels $\mathbf{H}_{\mathrm{IE}}\in \mathbb{C} ^{N_{\mathrm{r}}\times N_{\mathrm{t}}}$ can be obtained. 
Then, the initially estimated channels instead of the received signals are fed into the proposed IE-Pix2pix as the conditional input. Finally, the proposed IE-Pix2pix outputs the estimated near-field XL-MIMO channels through the mapping relationship
\begin{equation}
\hat{\mathbf{H}}=f_{\mathrm{GAN}}\left( \mathbf{H}_{\mathrm{IE}};\Phi _{\mathrm{GAN}} \right),
\end{equation}
where $f_{\mathrm{GAN}}$ denotes the GAN process  and $\Phi _{\mathrm{GAN}}$ represents the parameter set of the IE-Pix2pix.

\subsection{Objective Functions}

The channel estimation problem is treated as an image-to-image translation problem because the channel matrices can be treated as two-channel images. 
Therefore, we are able to exploit the GAN and its variants to address the problem of channel estimation.
To estimate the channel matrices, GANs are used to learn the mapping from the random noise $\mathbf{Z}$ to the true channel matrices.
The noise $\mathbf{Z}$ is usually sampled from a prior distribution $p_{Z}$ (e.g., uniform distribution and Gaussian distribution).
However, this mapping learned by GANs may be unstable and the learning capacity of GANs is usually not good \cite{b23, b24} for channel estimation tasks. 
To tackle this issue, both the random noise $\mathbf{Z}$ and the conditional input are input into the CGAN to generate channel matrices with a specific property.  
As one of the GAN variants, Pix2pix is utilized to learn the mapping from the conditional input to the true channel matrices, which is similar to the CGAN. 
However, no random noise $\mathbf{Z}$ is directly fed into the Pix2pix.
Although the Pix2pix without the noise $\mathbf{Z}$ as the input may produce deterministic outputs, it can still learn the mapping from the conditional input to the channel matrices \cite{b25}.  
Note that as the input of the generator in the GAN and its variants, the random noise $\mathbf{Z}$ is not the noise in the environment but the random noise sampled from a prior distribution, e.g., uniform distribution. In this paper, the proposed IE-Pix2pix based on the Pix2pix is developed to estimate the XL-MIMO near-field channels more accurately by using the initially estimated channel matrices as the conditional input. 
In the following, we take the GAN based near-field channel estimation as an example to describe the objective functions and structures of the GAN and its variants.


\begin{figure*}[htbp]  
	\begin{minipage}[htbp]{0.5\textwidth}
		\centering	
		\includegraphics[width=7cm]{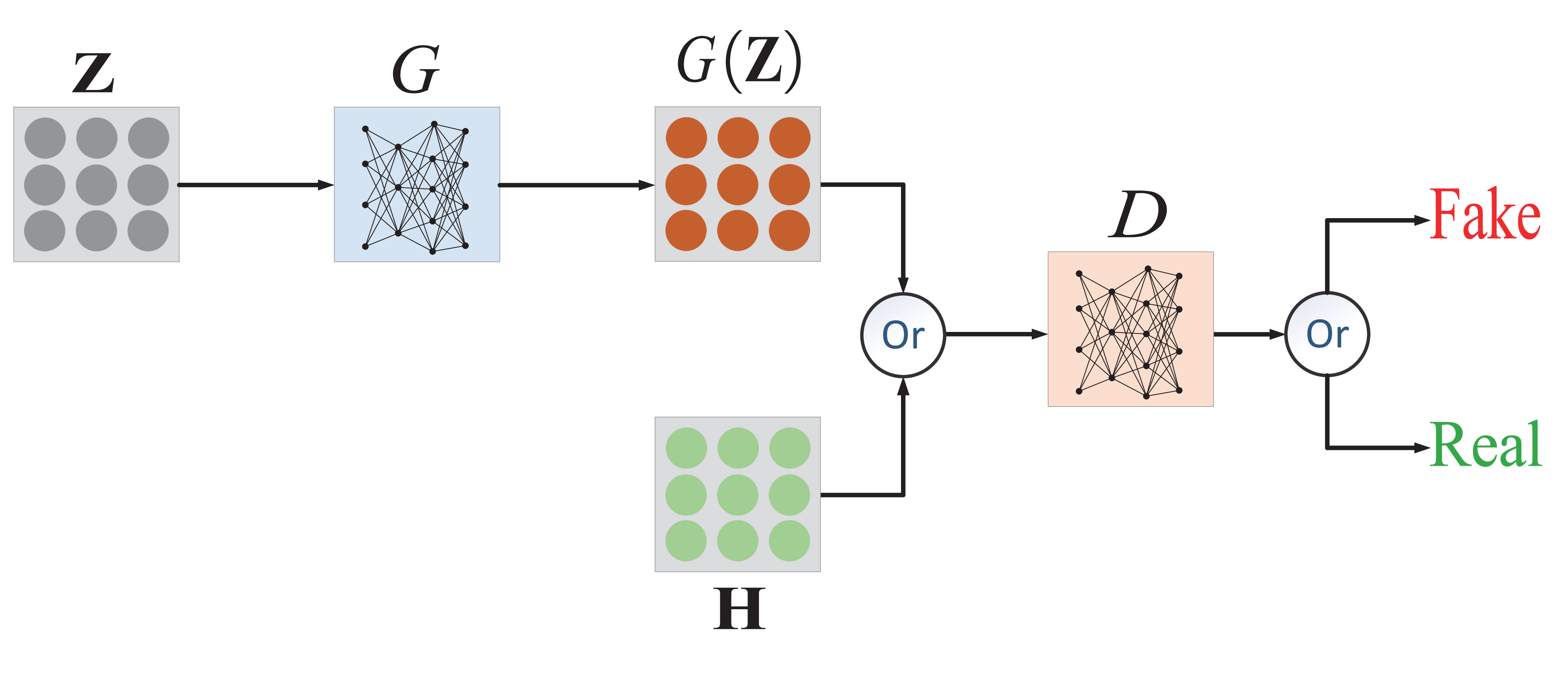}
		\subcaption{}
	\end{minipage}		
	\begin{minipage}[htbp]{0.3\textwidth}	
		\centering	
		\includegraphics[width=7cm]{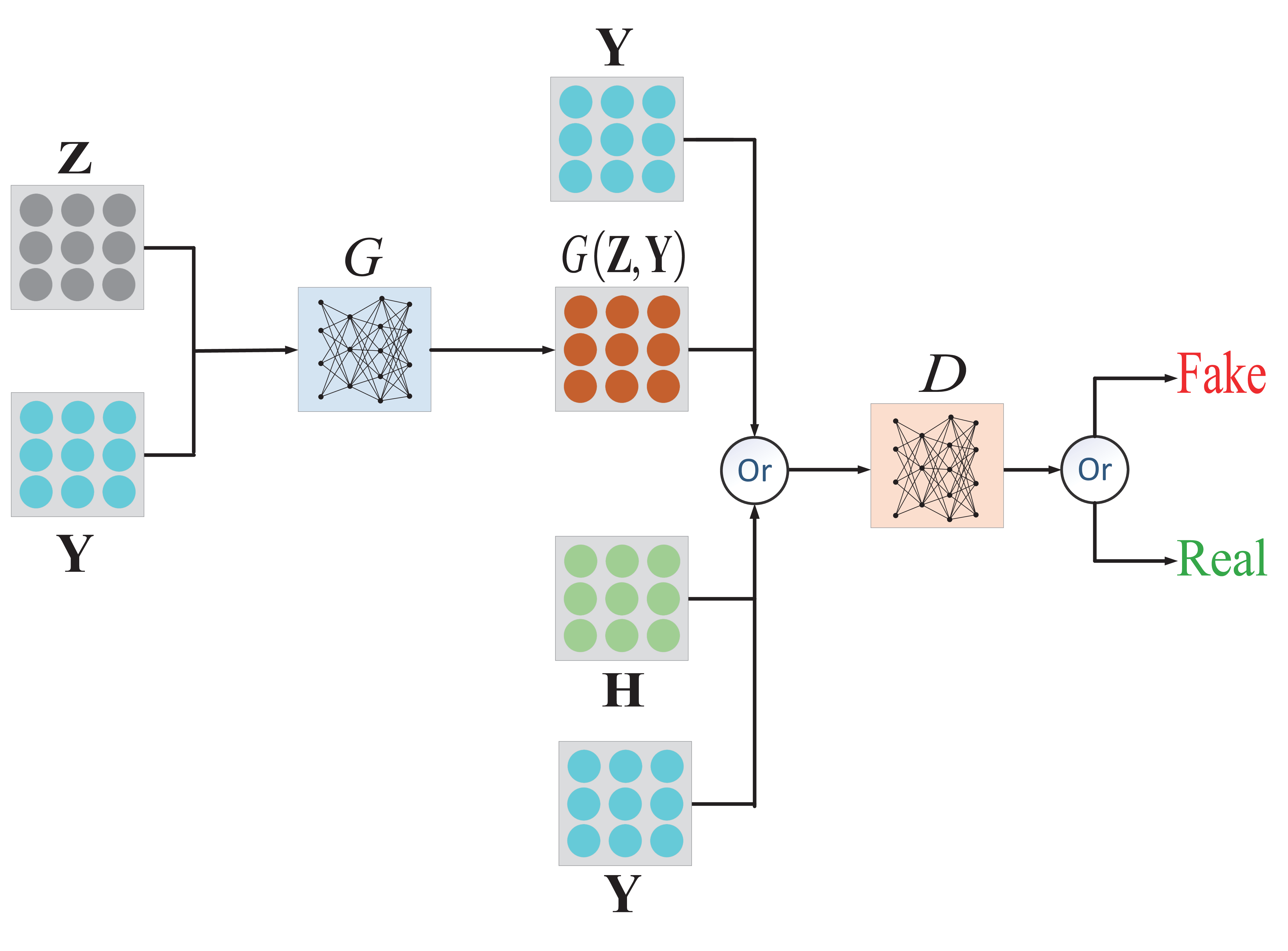}
		\subcaption{}
	\end{minipage}
	\caption{(a) Structure of GAN. (b) Structure of CGAN.}
\end{figure*}

Generally speaking, the GAN model is composed of two DL networks, i.e., the generator and the discriminator. 
For channel estimation tasks, the generator in the GAN model exploits the random noise to synthesize the mimic channel samples and fool the discriminator during the training phase, as shown in Fig. 3 (a). 
In contrast, the objective of the discriminator in the GAN model is to distinguish the mimic channel samples generated by the generator from the true channel samples conforming to the distribution of actual channel matrices $p_{H}$. 
Let $\varPhi_{G1}$ and $\varPhi_{D1}$ denote the parameter sets of the generator and the discriminator, respectively.
Furthermore, denote the generator and the discriminator as $G$ and $D$, respectively. 
Then, the objective functions for training the generator and the discriminator in the GAN model can be formulated as 
\begin{equation}
\mathcal{L}_{G\_\mathrm{GAN}}=\underset{\varPhi_{G1}}{\min}\,\left(\mathbb{E}_{\mathbf{Z}\sim p_Z}\left[ \log \left( 1-D\left( G\left( \mathbf{Z} \right) \right) \right) \right] \right),
\end{equation}
\begin{equation}
\begin{aligned}
\mathcal{L}_{D\_\mathrm{GAN}}=&\underset{\varPhi_{D1}}{\max}\left( \mathbb{E}_{\mathbf{H}\sim p_{H}}\left[ \log \left( D\left( \mathbf{H}\right) \right) \right] \right)\\
&+\mathbb{E}_{\mathbf{Z}\sim p_Z}\left[ \log \left( 1-D\left( G\left( \mathbf{Z} \right) \right) \right) \right],
\end{aligned}
\end{equation}
respectively. In the GAN model, $G\left(\mathbf{Z}\right)$ denotes the mimic channel samples generated by the generator, while $D\left(\mathbf{H}\right)$ represents the output of the discriminator. 
Generally speaking, the output of the discriminator is a real value between 0 and 1, which denotes the probability that the input belongs to the channel samples conforming to the distribution $p_H$.
Note that the output of the discriminator in the proposed IE-Pix2pix is a real-valued matrix and each of the elements in the matrix is a real value between 0 and 1. This is because the discriminator in the proposed IE-Pix2pix is with the patch architecture.
During the training process, the discriminator learns to maximize each element in the output of the discriminator as seen in (13). 
That is to say, the discriminator learns to make each element in $D\left(\mathbf{H}\right)$ approach 1. 
Meanwhile, the generator learns to minimize each element in $\log \left( 1-D\left( G\left( \mathbf{Z} \right) \right) \right)$, which implies that the generator learns to maximize each element in $D\left( G\left( \mathbf{Z} \right) \right)$. 
That is to say, the generator learns to make each element in $D\left(\mathbf{H}\right)$ approach 0. 
When the discriminator successfully distinguish the mimic channel samples generated by the generator from the true channel samples, it produces the feedback to the generator \cite{b27}. The generator is able to make the generated channel samples more accurate due to the feedback of the discriminator.
When the discriminator is not capable of distinguishing the mimic channel samples generated by the generator from the true channel samples, the training procedure of the GAN ends.    
Hence, the generator and the discriminator in the GAN model are adversarially trained to learn an adversarial loss during the learning phase.

As shown in Fig. 3 (b), we are able to extend the GAN to a CGAN by inputting the received signals $\mathbf{Y}$ into both the generator and the discriminator as the conditional input.
In the CGAN model, $G\left(\mathbf{Z},\mathbf{Y} \right)$ represents the mimic channel samples generated by the generator while the output of the discriminator is $ D\left(\mathbf{H},\mathbf{Y} \right)$ or $D\left( G\left( \mathbf{Z},\mathbf{Y} \right),\mathbf{Y} \right)$. 
In the CGAN model, the conditional input and the true channel samples or the mimic channel samples generated by the generator are concatenated and input into the discriminator. 
Let $\varPhi_{D2}$ and $\varPhi_{G2}$ be the parameter sets of the discriminator and the generator, respectively. 
The objective functions of training the generator and the discriminator are given by
\begin{equation}
\mathcal{L}_{G\_\mathrm{CGAN}}=\underset{\varPhi_{G2}}{\min}\,\left(\mathbb{E}_{\mathbf{Z}\sim p_Z}\left[ \log \left( 1-D\left( G\left( \mathbf{Z},\mathbf{Y} \right) ,\mathbf{Y} \right) \right) \right] \right),
\end{equation}
\begin{equation}
\begin{aligned}
\mathcal{L}_{D\_\mathrm{CGAN}}=&\underset{\varPhi_{D2}}{\max}\left( \mathbb{E}_{\mathbf{H}\sim p_{H}}\left[ \log \left( D\left( \mathbf{H},\mathbf{Y}\right) \right) \right] \right)\\
&+\mathbb{E}_{\mathbf{Z}\sim p_Z}\left[ \log \left( 1-D\left( G\left( \mathbf{Z},\mathbf{Y} \right),\mathbf{Y} \right) \right) \right],
\end{aligned}
\end{equation}
respectively.
In the CGAN model, the discriminator learns to maximize each of the elements in the output of the discriminator $D\left( \mathbf{H},\mathbf{Y}\right)$ as seen in (15). 
In the meanwhile, the generator tries to minimize each of the elements in $\log \left( 1-D\left( G\left( \mathbf{Z},\mathbf{Y} \right),\mathbf{Y} \right) \right)$ during the training phase, which means that the generator learns to make each of the elements in $D\left(\mathbf{H},\mathbf{Y} \right)$ approach 0.
During the training process, the generator and the discriminator in the CGAN model compete with each other to learn an adversarial loss. 
\begin{figure*}
	\centering
	\includegraphics[width=15cm]{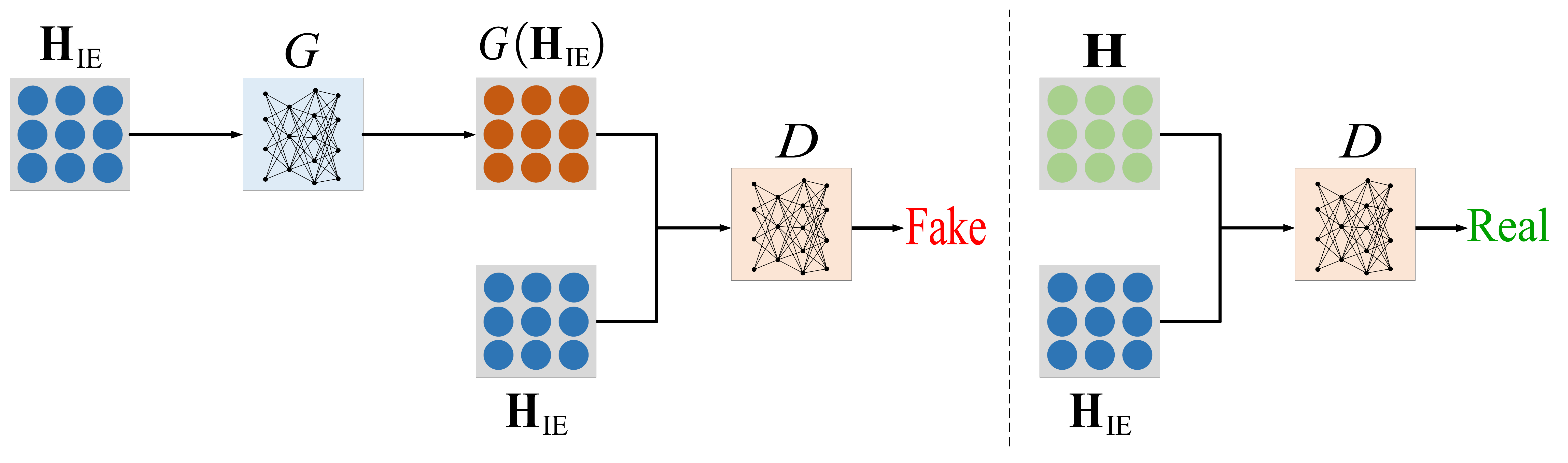}
	\caption{IE-Pix2pix based channel estimation.}
	\label{fig.3}
\end{figure*}

Similar to the CGAN model, the Pix2pix also learns the mapping from the conditional input to the channel samples for channel estimation tasks. However, no random noise is directly input into the generator in the Pix2pix. 
In the proposed IE-Pix2pix,  the generator learns the mapping from the initially estimated channels to the true channel matrices. 
As shown in Fig. 4, the initially estimated channel matrices instead of the received signals are input into both the generator and the discriminator in the proposed IE-Pix2pix as the conditional input.
Let $\varPhi_{D3}$ and $\varPhi_{G3}$ denote the parameter sets of the discriminator and the generator in the IE-Pix2pix, respectively. 
The objective functions for training the generator and the discriminator in the IE-Pix2pix are expressed as 
\begin{equation}
\mathcal{L}_{G\_\mathrm{Pix}}=\underset{\varPhi_{G3}}{\min}\,\left(\mathbb{E}\left[ \log \left( 1-D\left( G\left(\mathbf{H}_{\mathrm{IE}}   \right) ,\mathbf{H}_{\mathrm{IE}} \right) \right) \right] \right),
\end{equation}
\begin{equation}
\begin{aligned}
\mathcal{L}_{D\_\mathrm{Pix}}=&\underset{\varPhi_{D3}}{\max}\left( \mathbb{E}\left[ \log \left( D\left( \mathbf{H},\mathbf{H}_{\mathrm{IE}}\right) \right) \right] \right)\\
&+\mathbb{E}\left[ \log \left( 1-D\left( G\left(\mathbf{H}_{\mathrm{IE}} \right),\mathbf{H}_{\mathrm{IE}} \right) \right) \right],
\end{aligned}
\end{equation}
respectively. In the proposed IE-Pix2pix, $G\left(\mathbf{H}_{\mathrm{IE}}   \right)$ denotes the mimic channel samples produced by the generator, while $D\left( G\left(\mathbf{H}_{\mathrm{IE}} \right),\mathbf{H}_{\mathrm{IE}} \right)$ or $D\left( \mathbf{H},\mathbf{H}_{\mathrm{IE}} \right)$ represents the output of the discriminator.    
During the training, the generator learns to minimize each of the elements in $\log \left( 1-D\left( G\left(\mathbf{H}_{\mathrm{IE}} \right),\mathbf{H}_{\mathrm{IE}} \right) \right)$ as seen in (16), which denotes that the generator aims to maximize each of the elements in $D\left( G\left(\mathbf{H}_{\mathrm{IE}} \right),\mathbf{H}_{\mathrm{IE}} \right)$. That is to say, the generator learns to make each of the elements in $D\left( \mathbf{H},\mathbf{H}_{\mathrm{IE}} \right)$ approach 0. 
On the contrary, the objective of the discriminator is to make each of the elements in $D\left( \mathbf{H},\mathbf{H}_{\mathrm{IE}} \right)$ approach 1. 
Therefore, the generator and the discriminator are adversarially trained to learn an adversarial loss. The adversarial loss for the proposed IE-Pix2pix can be expressed as 
\begin{equation}
\begin{aligned}
\mathcal{L}_{\mathrm{Pix}}\left( G,D \right)=&\mathbb{E} \left[ \log \left( D\left( \mathbf{H},\mathbf{H}_{\mathrm{IE}} \right) \right) \right]\\
&+\mathbb{E}\left[ \log \left( 1-D\left( G\left(\mathbf{H}_{\mathrm{IE}} \right),\mathbf{H}_{\mathrm{IE}} \right) \right) \right].
\end{aligned}
\end{equation}

To ensure the correct direction of training the generator in the IE-Pix2pix, we mix the objective function for the generator with a conventional loss function \cite{b25}, which is given by 
\begin{equation}
\mathcal{L} _{\mathrm{L}1}\left( G \right) =\mathbb{E} \left[ \left\| \mathbf{H}-G\left(\mathbf{H}_{\mathrm{IE}} \right) \right\| \right].
\end{equation}

Moreover, we use a hyper-parameter $\eta$ to control the relative importance of the conventional loss function. 
We are able to further obtain performance improvement by adjusting the values of the hyper-parameter $\eta$.
Then, the objective function of training the generator in the IE-Pix2pix can be rewritten as  
\begin{equation}
\mathcal{L} _{G\_\mathrm{Pix}}=\underset{\varPhi _{G3}}{\min}\,\left( \mathbb{E} \left[ \log \left( 1-D\left( G\left( \mathbf{H}_{\mathrm{IE}} \right) ,\mathbf{H}_{\mathrm{IE}} \right) \right) \right] +\eta\mathcal{L} _{\mathrm{L}1}\left( G \right) \right).
\end{equation}

To ensure the correct direction of training the generator,
	we combine the adversarial
	loss with the conventional loss function. Finally, the training process of the IE-Pix2pix can be expressed as 
\begin{equation}
G^*=\mathrm{arg}\,\underset{\varPhi _{G3}}{\min}\,\underset{\varPhi _{D3}}{\max}\left( \mathcal{L}_{\mathrm{Pix}}\left(G,D \right)+\eta\mathcal{L} _{\mathrm{L}1}\left( G \right) \right).
\end{equation}

The proposed IE-Pix2pix model can be used to generate more accurate channel samples by solving the min-max problem in equation (21). During the training, the parameters of the discriminator and the generator are updated by leveraging equation (17) and equation (20), respectively.
The parameters of the discriminator are fixed, when the adversarially trained generator is trained.

\begin{table*}[htbp] 
	\caption{Architecture of the Proposed IE-Pix2pix}
	\centering
	\scalebox{1}
	{ 
		\begin{tabular}{c|c|c|c|c|c|c}
			
			\toprule[0.8pt]
			Network & Modules  & Layer &   \makecell[c]{Number of  filters} &
			
			\makecell[c]{Size of  filters}   &   \makecell[c]{Activation   function}  &   Stride  \\  \hline 
			&  Input & 1\,\,$\left( \bar{\mathbf{H}}_{\mathrm{IE}}\in \mathbb{C} ^{N_{\mathrm{r}}\times N_{\mathrm{t}}\times 2} \right)$ & / & / & / & /   \\  \cline{2-7}
			\multirow{6}*{\centering \makecell[c]{IE-Pix2pix}   }&\multirow{5}*{Generator}     &  2\,\,$ \left( \mathrm{ConvT} \right)  $  & 2 & $ 3\times 3 $  & ReLU & $\left( 1,1 \right)$ \\  \cline{3-7}
				
&  & 3\,\,$ \left( \mathrm{Conv} \right)  $& 2 & $ 3\times 3 $  & LReLU & $\left( 1,1 \right)$ \\  \cline{3-7}	
 
 &  &  $ 4\sim8 $\,\,$ \left( \mathrm{Conv} \right) $  & 64 & $ 3\times 3 $ &  LReLU & $\left(2,2 \right)$ \\  \cline{3-7}

&  &  $ 9\sim12 $\,\,$ \left( \mathrm{ConvT} \right) $  & 64 & $ 3\times 3 $  & ReLU & $\left(2,2 \right)$ \\  \cline{3-7}

&     &$ 13$\,\,$ \left( \mathrm{ConvT} \right) $ & 2 & $ 3\times 3 $  & Tanh  & $\left(2,2 \right)$ \\  \cline{3-7}
			
\cline{2-6}
			&\multirow{3}*{Discriminator}     & $ 14\sim16 $\,\,$ \left( \mathrm{Conv} \right) $ & 64 & $ 3\times 3 $  & LReLU & $\left(2,2 \right)$ \\  \cline{3-7}
			&  & $ 17 $\,\, $\left( \mathrm{Conv} \right) $   & 64 & $ 3\times 3 $  & LReLU & $\left(2,1 \right)$ \\  \cline{3-7}
			&     &$ 18 $\,\, $\left( \mathrm{Conv} \right) $ & 1 & $ 3\times 3 $  & Sigmoid & $\left(1,1 \right)$ \\   
			
			\bottomrule[0.8pt]
		\end{tabular}	
	}	
\end{table*}

\subsection{Network Architecture}

The proposed IE-Pix2pix contains the generator and the discriminator. Both the generator and the discriminator are composed of convolutional layers. 
For the proposed IE-Pix2pix, 
we adopt the U-Net architecture and the patch architecture \cite{b23, b25} in the generator and the discriminator, respectively. 
The generator in the IE-Pix2pix  mainly includes five encoder blocks and four decoder blocks. Each of the encoder blocks contains a convolutional layer (called Conv), a batch normalization layer,  and a leaky-rectified linear unit (LReLU) activation layer. Every decoder block consists of a transposed convolutional layer (called ConvT), a batch normalization layer, and a rectified linear unit (ReLU) activation layer.
In addition, one transposed convolutional layer and one convolutional layer are used as the first two layers of the generator, which can rescale the input of the generator into the same size as $\mathbf{H}$. 
When the conditional input and the near-field channel matrix are of different sizes, the conditional input can be rescaled into the same size as the near-field channel matrix by changing the strides in the first two layers of the generator. 
Note that a transposed convolutional layer and a convolutional layer can be treated as a upsampling layer and a downsampling layer, respectively. 
Filters with the size of $3\times3$ are used for all convolutional layers. 
Different from the common generator, the generator in the IE-Pix2pix is an encoder-decoder CNN with skip connections.  
The pixel level details can be preserved by exploiting the skip connections to combine the feature maps of the decoder blocks and the encoder blocks. 
The tangent activation function is used as the last layer of the generator to scale the output values of the generator into the 
$\left[ -1,1 \right]$ range, which is able to avoid the mode collapse. 
The detailed architecture of the IE-Pix2pix is shown in Table I. 
Let stride = $\left( s^{\mathrm{h}},s^{\mathrm{v}} \right) $ in Tensorflow, where $s^{\mathrm{v}}$ and $s^{\mathrm{h}}$ 
	are the vertical and horizontal strides, respectively. Let $F_{s-1,w}$ and $F_{s-1,h}$ denote the width and
	the height of the input feature map in the $s$-th layer, respectively. The width and the height of the output feature map in the $s$-th transposed convolutional layer can be expressed as 
	$F_{s,w}=\lceil F_{s-1,w}\times s^{\mathrm{h}} \rceil$ and $F_{s,h}=\lceil F_{s-1,h}\times s^{\mathrm{v}} \rceil$, respectively. The width and
	the height of the output feature map in the $s$-th convolutional layer can be expressed as 
	$F_{s,w}=\lceil \frac{F_{s-1,w}}{s^{\mathrm{h}}} \rceil$ and $F_{s,h}=\lceil \frac{F_{s-1,h}}{s^{\mathrm{v}}} \rceil$, respectively.

Different from the common discriminator, the patch discriminator adopted in this paper is capable of mapping the input to a receptive field instead of a scalar output. 
Each of elements in the receptive field represents whether the input of the discriminator is real or fake. Then, the final output of the discriminator can be obtained by averaging all responses of the receptive field.  
Therefore, the patch discriminator is able to recognize the information of the input more efficiently. 
The patch discriminator is mainly composed of four encoder blocks. To acquire the receptive field, we choose a convolutional layer instead of a fully connected layer as the last layer of the discriminator. Furthermore, the Sigmoid activation function is applied for the final layer of the discriminator, which can scale the output values of the discriminator to the $\left[0,1 \right]$ range.

\subsection{IE-Pix2pix Based Channel Estimation}

In order to estimate the XL-MIMO near-field channels more accurately, we exploit the idea of GANs to propose a channel estimation method based on the IE-Pix2pix. Specifically, a GAN variant (i.e., Pix2pix) is designed for channel estimation in XL-MIMO systems. Moreover, we combine the objective function of the Pix2pix with a traditional loss function, which can ensure the appropriate direction of training the generator. The performance improvement can be obtained by changing the values of the hyper-parameter of the loss function. Additionally, the initially estimated channels instead of the received signals are input into the Pix2pix as the conditional input to estimate the XL-MIMO near-field channels more efficiently.

Next, we discuss how the proposed IE-Pix2pix model works for the near-field channel estimation problem in XL-MIMO systems. 
Like the vast majority of existing DL based schemes, the IE-Pix2pix mainly operates in two phases, namely, offline training phase and online prediction phase. 
Given a vast number of known training samples, the IE-Pix2pix aims to solve the min-max problem in equation (21) in the offline training phase. 
In the online prediction phase, the trained generator in the IE-Pix2pix can output the estimated near-field channels $\hat{\mathbf{H}}$ by employing the new input of $\mathbf{H}_{\mathrm{IE}}$.

\begin{algorithm}[h]  
 
	\caption{IE-Pix2pix Based Channel Estimation} 
	
\begin{algorithmic}[1] 

\Require  The received signals $\mathbf{Y}$ and the true near-field channel matrices $\mathbf{H}$.

\Ensure Estimated near-field channel matrices $\mathbf{\hat{H}}$.


\State Obtain the initially estimated near-field channels $\mathbf{H}_{\mathrm{IE}}$ by exploiting equation (10).

\For{the number of training iterations}
		
\State // Updating the generator;

\State  Sample minibatch of the initially estimated near-field channels $\mathbf{H}_{\mathrm{IE}}$.

\State Sample minibatch of the true near-field channels $\mathbf{H}$.

\State Feed the initially estimated near-field channels $\mathbf{H}_{\mathrm{IE}}$ into the proposed IE-Pix2pix as the conditional input.

\State 	Combine the conventional loss function of equation (19) with the objective function of the IE-Pix2pix in equation (18).

\State Update the parameters of the generator by solving the min-max problem in equation (21) for the IE-Pix2pix.
			
\State // Updating the discriminator; 
	
\State Sample minibatch of the initially estimated near-field channels $\mathbf{H}_{\mathrm{IE}}$.

\State Sample minibatch of the true near-field channels $\mathbf{H}$.

\State Update the parameters of the discriminator by solving the min-max problem of equation (21) for the IE-Pix2pix.
		
\EndFor

\end{algorithmic} 
	\Return The finally estimated near-field channels $ \mathbf{\hat{H}}$. 
	
\end{algorithm}

\textit{1) Offline training phase:}  
According to the adopted mixed LoS/NLoS XL-MIMO near-field channel model \cite{b16}, the training set composed of $I$ samples is generated to train and test the proposed IE-Pix2pix model. 
In our study, we adopt the supervised learning to train the generator in the IE-Pix2pix model.
The training set can be expressed as $\left\{ \mathbf{H}_{\mathrm{IE}}^{i},\mathbf{H}^i \right\} _{i=1}^{I}$, where $\mathbf{H}_{\mathrm{IE}}$ denotes the input of the generator, $\mathbf{H}$ represents the corresponding label, and $\left\{ \mathbf{H}_{\mathrm{IE}}^{i},\mathbf{H}^i \right\} $ is the $ i $-th sample.  
Note that both the real and imaginary parts of $\mathbf{H}_{\mathrm{IE}}$ and $\mathbf{H}$ are concatenated in the third dimension and input into the IE-Pix2pix, since the neural network software utilized in this work can not support complex-valued operations.   
To train and test the proposed IE-Pix2pix, the real and imaginary parts of $\mathbf{H}_{\mathrm{IE}}$ and $\mathbf{H}$ are concatenated in the third dimension to obtain $\bar{\mathbf{H}}_{\mathrm{IE}}^{i}\in \mathbb{C} ^{N_{\mathrm{r}}\times N_{\mathrm{t}}\times 2}$ and $\bar{\mathbf{H}}^i\in \mathbb{C} ^{N_{\mathrm{r}}\times N_{\mathrm{t}}\times 2}$, respectively.

We utilize $N_{\mathrm{t}}=256,N_{\mathrm{r}}=128$ as a typical example to describe the IE-Pix2pix based channel estimation for XL-MIMO systems. 
As shown in Fig. 4, the generator in the IE-Pix2pix receives the initially estimated complex-valued channel matrix $\mathbf{H}_{\mathrm{IE}}^{i}\in \mathbb{C} ^{128\times 256}$ as conditional input to produce mimic complex-valued channel matrix $\hat{\mathbf{H}}^i\in \mathbb{C} ^{128\times 256}$ (i.e., $G\left( \mathbf{H}_{\mathrm{IE}}^{i}\in \mathbb{C} ^{128\times 256} \right)$).  
In fact, the real and imaginary parts of $\mathbf{H}_{\mathrm{IE}}^{i}\in \mathbb{C} ^{128\times 256}$ and $\mathbf{H}^i\in \mathbb{C} ^{128\times 256}$ are concatenated in the third dimension to obtain the real-valued matrices 
$\bar{\mathbf{H}}_{\mathrm{IE}}^{i}\in \mathbb{C} ^{128\times 256\times 2}$ and $ \bar{\mathbf{H}}^i\in \mathbb{C} ^{128\times 256\times 2} $, respectively.  
Then, two real-valued matrices
$\bar{\mathbf{H}}_{\mathrm{IE}}^{i}$ are input into the generator to generate two estimated real-valued channel samples  $\hat{\bar{\mathbf{H}}}^i\in \mathbb{C} ^{128\times 256\times 2}$ (i.e., $G\left( \bar{\mathbf{H}}_{\mathrm{IE}}^{i}\in \mathbb{C} ^{128\times 256\times 2} \right)$).
Specifically, the two real-valued matrices 
$\bar{\mathbf{H}}_{\mathrm{IE}}^{i}$ are processed by the first two layers of the generator. The two feature matrices $\hat{\bar{\mathbf{H}}}^i\in \mathbb{C} ^{128\times 256\times 2}$ have the same size as $\bar{\mathbf{H}}_{\mathrm{IE}}^{i}$. This is because the stride in the first two layers of the generator is set as 1 when the conditional input and the near-field channel are of the same size in the IE-Pix2pix. Then, the two feature matrices are processed by five encoder blocks and four decoder blocks in the generator to generate 64 real-valued feature matrices. After that, the 64 real-valued feature matrices are further processed by the last layer of the generator to produce 2 $128\times 256$ estimated real-valued channel matrices, i.e., $\hat{\bar{\mathbf{H}}}^i\in \mathbb{C} ^{128\times 256\times 2}$ ($ G\left( \bar{\mathbf{H}}_{\mathrm{IE}}^{i} \right) $). 
Finally, the estimated real-valued channel matrices $\hat{\bar{\mathbf{H}}}^i\in \mathbb{C} ^{128\times 256\times 2}$ or the true real-valued channel matrices $\bar{\mathbf{H}}^i\in \mathbb{C} ^{128\times 256\times 2}$ and the real-valued matrices 
$\bar{\mathbf{H}}_{\mathrm{IE}}^{i}\in \mathbb{C} ^{128\times 256\times 2}$ are concatenated and then fed into the discriminator. 
For the discriminator, the real-valued matrices are processed by four encoder blocks in the discriminator to generate 64 real-valued feature matrices. Then, the 64 real-valued feature matrices are processed by the final layer of the discriminator to generate one real-valued matrix.
Each element in the real-valued matrix is a real value between 0 and 1, which denotes whether the input of the discriminator is real or fake.
The final output of the discriminator is obtained by averaging all the elements in the real-valued matrix. Note that the Sigmoid activation function is used to scale the discriminator's output values to the [0, 1] range.

When the input of the discriminator contains the estimated real-valued channel matrices   $\hat{\bar{\mathbf{H}}}^i\in \mathbb{C} ^{128\times 256\times 2}$, the discriminator learns to output a low value and recognize its input as a fake label, as shown in Fig. 4. 
In the meanwhile, the discriminator aims to output a high value and recognize its input as a real label, when the input of the discriminator includes the true real-valued channel matrices   $\bar{\mathbf{H}}^i\in \mathbb{C} ^{128\times 256\times 2}$. 
Owing to the feedback of the discriminator, the generator is capable of enhancing the ability of generating mimic channel samples similar to the true channel samples during the training process.    
The $128\times 256$ complex-valued  estimated channel matrix $\hat{\mathbf{H}}^i\in \mathbb{C} ^{128\times 256}$ can be obtained by combining the two estimated real-valued channel matrices $\hat{\bar{\mathbf{H}}}^i\in \mathbb{C} ^{128\times 256\times 2}$.
In the offline training phase, the objective of the IE-Pix2pix is to solve the min-max problem of  equation (21). 
During the training, the parameters of the discriminator and the generator are updated by employing equation (17) and equation (20), respectively.  
The specific training procedure for the proposed IE-Pix2pix is shown in Algorithm 1.  The parameters of the discriminator are fixed, while the adversarially trained generator is trained.


\textit{2) Online prediction phase:}
In the online prediction phase, the trained generator in the IE-Pix2pix model is applied to the near-field channel estimation problem in XL-MIMO systems. After the offline training phase, we can obtain the trained generator in the IE-Pix2pix. Then, the trained generator can be used to perform near-field channel
estimation by using the new initially estimated near-field channels $\mathbf{H}_{\mathrm{IE}}$ in the online prediction phase.
The new initially estimated near-field channels $\mathbf{H}_{\mathrm{IE}}$ are input into the trained generator in the IE-Pix2pix to output the estimated near-field channels $\hat{\mathbf{H}}$ composed of both the LoS and NLoS path components.

Eventually, we employ the normalized mean-squared error (NMSE) to evaluate the channel estimation performance of the IE-Pix2pix model:

 \begin{equation}
\mathrm{NMSE}=\frac{\mathbb{E}\left[ \left\| \mathbf{H} - \mathbf{\hat{H}} \right\| _{2}^{2} \right]}{\mathbb{E}\left[ \left\| \mathbf{H} \right\| _{2}^{2} \right]}.
\end{equation}

\subsection{Complexity Analysis}

In this subsection, we provide the complexity analysis of the proposed IE-Pix2pix based channel estimation method and existing channel estimation schemes. 
The computational complexity of the far-field codebook based OMP algorithm can be represented by $\mathcal{O} \left( N_{\mathrm{t}}N_{\mathrm{r}}\left( N_{\mathrm{t}}+N_{\mathrm{r}} \right) L \right)$ \cite{bb10}.
Let $S_{\mathrm{t}}$ and $S_{\mathrm{r}}$ denote the numbers of columns of the corresponding two polar-domain transform matrices, respectively.
Then, the computational complexity of the two stage channel estimation approach proposed in \cite{b16} can be presented as 
\begin{equation}
\mathcal{O} \left( \left( S_{\mathrm{LoS}}+P\varGamma \right) M_{\mathrm{r}}N_{\mathrm{t}}N_{\mathrm{r}}+N_{\mathrm{t}}N_{\mathrm{r}}\left( S_{\mathrm{t}}+S_{\mathrm{r}} \right)L \right),
\end{equation}
where $S_{\mathrm{LoS}}$ and $\varGamma$ denote the size of the parameters collection and the number of iteration, respectively.

The computational complexity of the proposed IE-Pix2pix comes from two parts, i.e., initial estimation in equation (10) and the designed Pix2pix. 
The complexity of the initial estimation in equation (10) can be represented by $N_{\mathrm{t}}N_{\mathrm{r}}\left( N_{\mathrm{t}}+N_{\mathrm{r}} \right)$.
The complexity of the designed Pix2pix comes from two CNNs, i.e., the generator and the discriminator. 
According to \cite{b28}, the complexity of the Pix2pix can be expressed as $\mathcal{O} \left( \sum\nolimits_{s=1}^S{F_{s,h}F_{s,w}K_{s}^{2}N_sN_{s-1}} \right)$, where $ K_s $ and  $N_s$ represent the filter size and the number of output feature map in the $ s $-th layer, respectively.
According to the aforementioned offline training of the IE-Pix2pix, the related parameters are listed in Table II, when $N_{\mathrm{t}}=256$ and $N_{\mathrm{r}}=128$. Therefore, the computational complexity of the proposed IE-Pix2pix can be expressed as 
\begin{equation}
\mathcal{O} \left( N_{\mathrm{t}}N_{\mathrm{r}}\left( N_{\mathrm{t}}+N_{\mathrm{r}} \right) +\sum\nolimits_{s=1}^S{F_{s,h}F_{s,w}K_{s}^{2}N_sN_{s-1}} \right).
\end{equation}

\begin{table}[t]
	\centering
	\centering
	\caption{The Parameter Settings of the IE-Pix2pix}
	\scalebox{1}
	{
		\begin{tabular}{c|c|c|c|c|c}
			\toprule[0.8pt]
			$ s $  & $ F_{s,h} $ & $ F_{s,w} $ & $ K_{s} $  & $ N_{s} $ & $ N_{s-1} $    \\   \hline  
			$ 2 $  & 128 & 256 & 3 & 2  & 2    \\  \hline  
			$ 3 $  & 128 & 256 & 3 & 2 & 2     \\  \hline     
			$ 4 $  & 64 & 128 & 3 & 64  & 2   \\  \hline  		
			$ 5 $  & 32 & 64 & 3 & 64  & 64   \\  \hline  
			$ 6 $  & 16 & 32 & 3 & 64  & 64     \\  \hline  	
			$ 7 $  & 8 &  16 & 3 & 64  & 64    \\  \hline  		
			$ 8 $  & 4 & 8 & 3 & 64  & 64     \\  \hline  
			$ 9 $  & 8 & 16 & 3 & 64  & 64      \\  \hline 
			9  & 16 & 32 & 3  & 64 & 64    \\  \hline  
			$ 11 $  & 32 & 64 & 3 & 64   & 64 \\  \hline 
			$ 12 $  & 64 & 128 & 3 & 64  & 64  \\  \hline 
			$ 13 $  & 128 & 256 & 3 & 2 & 64   \\  \hline 
			$ 14 $  & 64 & 128 & 3 & 64 & 64   \\  \hline 
			$ 15 $  & 32 & 64 & 3 & 64  & 64  \\  \hline 	
			$ 16 $  & 16 & 32 & 3 & 64 & 64     \\  
			\hline 
			$ 17 $  & 8 & 32 & 3 & 64 & 64    \\ \hline 		
			$ 18 $  & 8 & 32 & 3 & 1 & 64     \\

			\bottomrule[0.8pt]
		\end{tabular}	
	}
 
\end{table}

\begin{table}[ht]
	\centering
	
	\caption{The Impact of $\eta$ on the NMSE Performance of the IE-Pix2pix}
	
	\scalebox{1}
	{
		\begin{tabular}{c|c|c}
			
			\toprule[0.8pt]
			$ \eta  $ &  NMSE (40 m)  &   NMSE (60 m)     \\   \hline  
			
			0.1 & \textbf{-8.784} & \textbf{-5.986}   \\   
			
			1 &-8.941& \textbf{-5.122}  \\    	   
			
			10 & -8.965 & -6.071  \\   
			
			100 & \textbf{-8.996} & \textbf{-6.091}   \\   	 		
			1000 & -8.959 & -6.071    \\   
			\bottomrule[0.8pt]
		\end{tabular}	
	}   
\end{table}

\section{Numerical Results}

In this section, we present the XL-MIMO near-field channel estimation performance comparison among the proposed IE-Pix2pix, the two stage algorithm proposed for the adopted channel model \cite{b16}, the CRLB, and the far-field codebook based OMP scheme \cite{bb10}.


\subsection{Simulation Setting}  

We consider the XL-MIMO near-field system, where the transmitter is equipped with $N_{\mathrm{t}}=256$ antennas and the receiver is equipped with $N_{\mathrm{r}}=128$ antennas. In this paper, we adopt the mixed LoS/NLoS XL-MIMO near-field channel model proposed in \cite{b16} and mainly focus on the channel estimation in the mixed LoS/NLoS near-field region. In this case, the distance between the transmitter and the receiver is smaller than the ARD, which can be calculated as $\mathrm{ARD}=\frac{4D_{\mathrm{t}}D_{\mathrm{r}}}{\lambda}\approx 197$ m.  
The carrier frequency is set as $f_{\mathrm{c}}=50$ GHz, which corresponds to $\lambda =0.006$ m.  
The XL-MIMO near-field channel in equation (3) is composed of $L=3$ NLoS path components. 
The angle $\theta _{\mathrm{t},l}$, angle $\theta _{\mathrm{r},l}$, and complex gain $\alpha_l$ are generated as following: $\theta _{\mathrm{t},l}\sim \mathcal{U} \left( -\frac{\pi}{3},\frac{\pi}{3} \right)$, $\theta _{\mathrm{r},l}\sim \mathcal{U} \left( -\frac{\pi}{3},\frac{\pi}{3} \right)$, and $\alpha _l\sim \mathcal{C} \mathcal{N} \left( 0,1 \right)$.  
Similar to \cite{b16}, both the distance $d_{l}^{\mathrm{t}}$ and distance $d_{l}^{\mathrm{r}}$ are generated in range of $ [10, 130] $ meters.
Each element in the combining matrix $ \mathbf{W} $  and the pilot matrix $ \mathbf{Q} $ is randomly selected from $\left\{ -\frac{1}{\sqrt{N_{\mathrm{r}}}},+\frac{1}{\sqrt{N_{\mathrm{r}}}} \right\} $ and $\left\{ -\frac{1}{\sqrt{P}},+\frac{1}{\sqrt{P}} \right\}$, respectively.

According to the XL-MIMO near-field channel model in equation (9), we generate channel samples to train and test the proposed IE-Pix2pix.   
Once the channel samples are generated, the corresponding received signals and initially estimated channel samples can be obtained by using equations (2) and (10), respectively.
The training set and test set consist of 4000 and 1000 samples, respectively. The proposed IE-Pix2pix is implemented in Python 3.7.16 with Tensorflow 2.2.0.
The optimizer for the generator and the discriminator is RMSProp. The learning rates for these two CNNs are 0.00002 and 0.0002, respectively.
The batch size is 1.

\begin{figure}
	\centering
	\includegraphics[width=7.5cm]{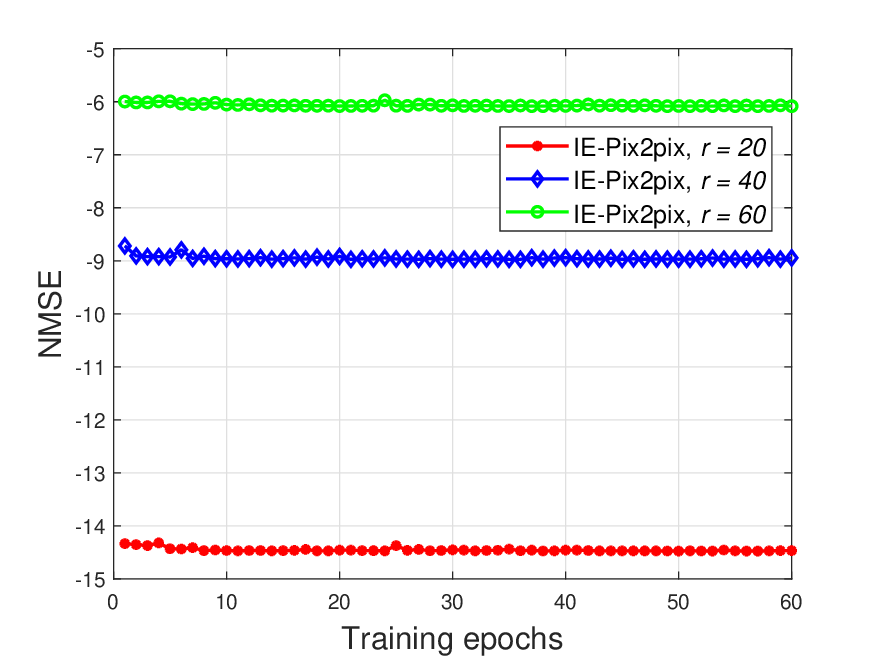}
	\caption{Training process of the proposed IE-Pix2pix.}
	\label{fig.2}
\end{figure}

Next, we investigate the impact of the hyper-parameter $\eta$ on the channel estimation performance of the IE-Pix2pix.
As shown in Table III, we are able to obtain the performance improvement by adjusting the values of the hyper-parameter $\eta$.   
It can be observed that the IE-Pix2pix achieves the best NMSE performance in Table III when SNR $=10$ dB.
Consequently, we empirically set $\eta =100$ in equation (21) for the following results unless stated otherwise.

\begin{figure}
	\centering
	\includegraphics[width=7.5cm]{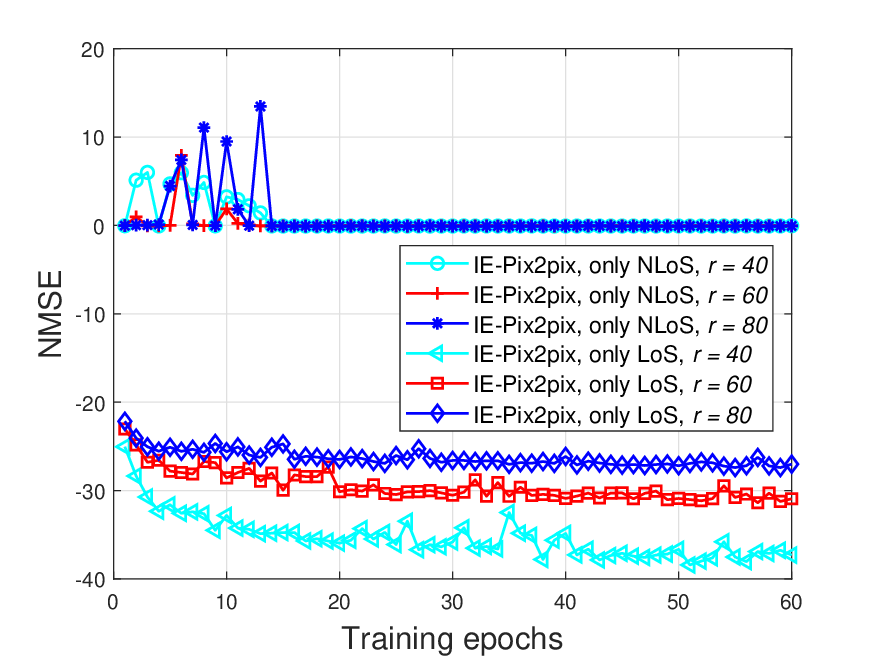}
	\caption{Training process of the IE-Pix2pix to only estimate the LoS path component or NLoS path components.}
	\label{fig.6}
\end{figure}

\subsection{IE-Pix2pix Based Channel Estimation}

\begin{figure}
	\centering
	\includegraphics[width=7.5cm]{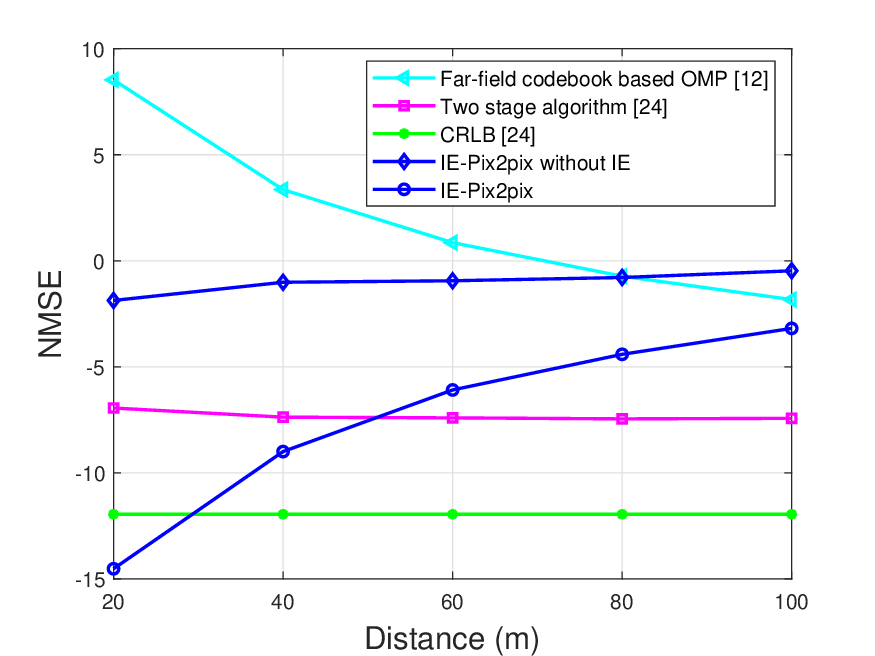}  
	\caption{NMSE performance against the distance for different methods.}
	\label{fig.6}
\end{figure}

Fig. 5 depicts the NMSE performance comparison of the proposed IE-Pix2pix under different distances. 
In Fig. 5, we train and test the proposed IE-Pix2pix under SNR $=10$ dB.
It can be observed that the NMSE performance of the IE-Pix2pix significantly degrades as the distance $  r$ increases.
This is due to the fact that the XL-MIMO near-field channel's LoS path component has different features when the distance between the transmitter and the receiver varies.   
Moreover, the bad estimation performance of the IE-Pix2pix mainly results from the large estimation error of the NLoS path components, which can be propagated to the estimation of the XL-MIMO channel containing the LoS and NLoS path components.    
In addition, the IE-Pix2pix with different distances converges very quickly.


\begin{figure*}[htbp]  
	\begin{minipage}[t]{0.5\textwidth}
		\centering	
		\includegraphics[width=7.5cm]{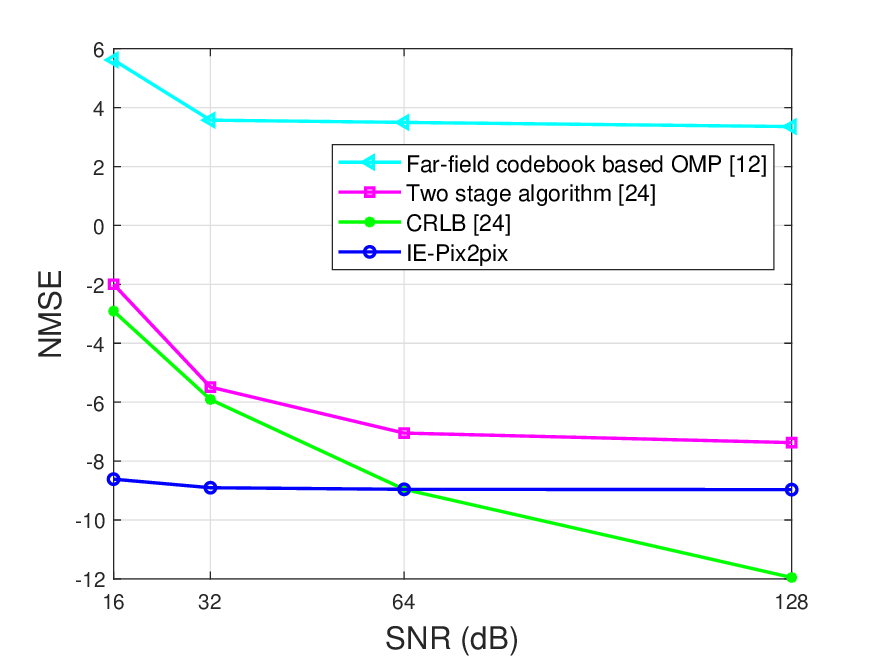} 
		\subcaption{}
	\end{minipage}		
	\begin{minipage}[t]{0.6\textwidth}			
		\centering	
		\includegraphics[width=7.5cm]{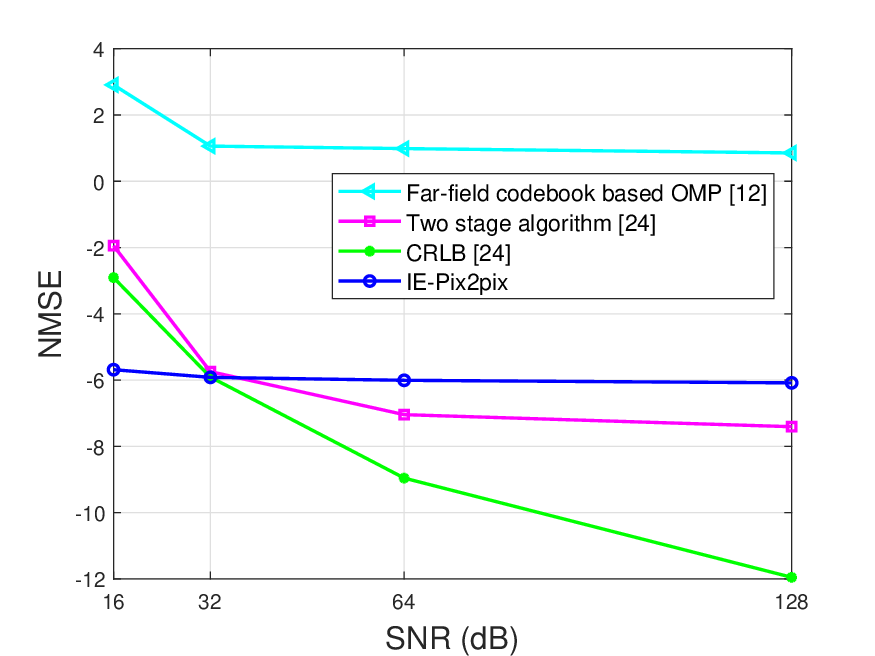} 
		\subcaption{}
	\end{minipage}	
	
	\caption{NMSE Performance comparison under different sizes of the pilot.
		(a) $r=40$.	(b) $r=60$.}	
\end{figure*}

We also test the NMSE performance of the proposed IE-Pix2pix under SNR $=10$ dB when the IE-Pix2pix is used to estimate the XL-MIMO near-field channel only including the LoS path component or NLoS path components.
Fig. 6 shows the training process of the IE-Pix2pix to only estimate the LoS path component or NLoS path components. When the XL-MIMO near-field channel only includes the NLoS path components, the proposed IE-Pix2pix achieves very bad NMSE performance.
However,  the IE-Pix2pix has very good NMSE performance when the IE-Pix2pix is used to estimate the XL-MIMO near-field channel only containing the LoS path component.
Furthermore, the NMSE performance of the IE-Pix2pix to only estimate the LoS path component improves with the decrease in the distance $r$.
This is because the LoS path component of the XL-MIMO near-field channel has different features with the change of the distance $r$ \cite{b16}.
Hence, 
the NMSE performance
of the IE-Pix2pix to estimate the XL-MIMO near-field channel including both the LoS and NLoS path
components degrades gradually with the increase in the distance $r$, as shown in Fig. 5 and Fig. 7.
The bad performance of
the IE-Pix2pix to simultaneously estimate the LoS
and NLoS path components is mainly due to the large estimation error of
the NLoS path components.
Note that the objective of the proposed IE-Pix2pix is to simultaneously estimate the LoS and NLoS path components. 
Although the IE-Pix2pix has very bad performance when the IE-Pix2pix is used to estimate the XL-MIMO channel only including the NLoS path components, the IE-Pix2pix can still achieve satisfactory NMSE performance in the adopted mixed LoS/NLoS channel model when the distance $r$ is small. 
This is because the IE-Pix2pix can still efficiently learn the mapping from the initially estimated channels to the XL-MIMO near-field channels composed of the LoS and NLoS path components.

Fig. 7 shows the impact of the distance $r$ on the NMSE performance when SNR $=$ 10 dB.
The size of the pilot matrix is $256\times 128$.
As the distance $r$ decreases, the NMSE performance of the proposed IE-Pix2pix improves gradually while the NMSE performance of the far-field codebook based OMP scheme degrades. 
This is because there are phase discrepancies between the far-field array response vectors based channel and the adopted channel \cite{b16}, which can be learned by the IE-Pix2pix efficiently.
The proposed IE-Pix2pix achieves better NMSE performance than the two stage algorithm when $ r\leqslant $ 50 m. 
The IE-Pix2pix without the IE can still outperform the far-field codebook based OMP method when $ r\leqslant $ 80 m, while the IE-Pix2pix performs much better than the IE-Pix2pix without the IE. 
This result demonstrates that the initially estimated channels can help the proposed IE-Pix2pix generate the XL-MIMO near-field channels more efficiently.
For the IE-Pix2pix without the IE,   
the strides in the first two layers of the generator need to be changed to rescale the received signals $\mathbf{Y}$ into the same size as the XL-MIMO near-field channels $\mathbf{H}$, since $\mathbf{Y}$ and $\mathbf{H}$ are of different sizes.
Furthermore, the IE-Pix2pix can even outperform the CRLB when the distance $r$ is enough small.
In addition, both the CRLB and the two stage algorithm achieve almost unchanged NMSE performance when the distance $r$ varies.
Note that the CRLB is the best unbiased estimator but not the greatest lower bound.

Next, we show the NMSE performance with various sizes of the pilot when SNR $=10$ dB in Fig. 8.
It is observed from Fig. 8 (a) that the NMSE performance of the IE-Pix2pix degrades slightly with the decrease of the pilot length. 
This phenomenon means that the IE-Pix2pix can still generate the XL-MIMO near-field channels efficiently according to the initially estimated channels when the pilot length decreases.
The initially estimated channel still has the same size as the XL-MIMO near-field channel when the pilot overhead is low. Hence, the IE-Pix2pix is able to generate the XL-MIMO channels more efficiently. 
Furthermore, the adversarial loss introduced by the GAN architecture helps the IE-Pix2pix adapt to different noise of the XL-MIMO channels. 
On the contrary, the NMSE performance of the two stage algorithm and the CRLB significantly degrades as the pilot length decreases. 
The IE-Pix2pix has much better performance than the two stage algorithm when the distance $r=$ 40.
In addition, the proposed method outperforms the CRLB when the pilot length is smaller than 64.
Therefore, the IE-Pix2pix performs better at low pilot overhead.

\begin{figure*}[htbp]  
	
	\begin{minipage}[t]{0.5\textwidth}
		\centering	
		\includegraphics[width=7.5cm]{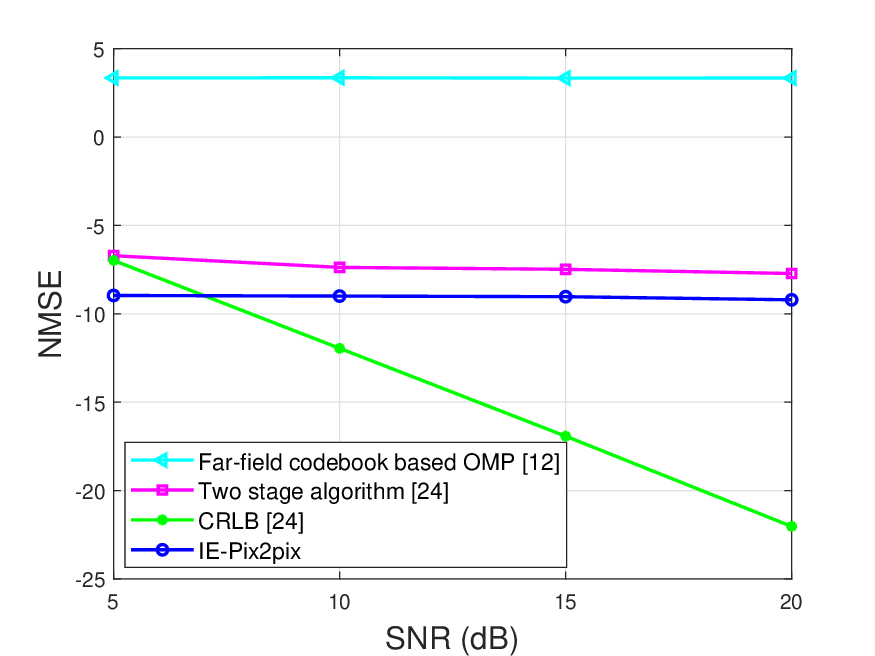}
		\subcaption{}
	\end{minipage}		
	\begin{minipage}[t]{0.6\textwidth}			
		\centering	
		\includegraphics[width=7.5cm]{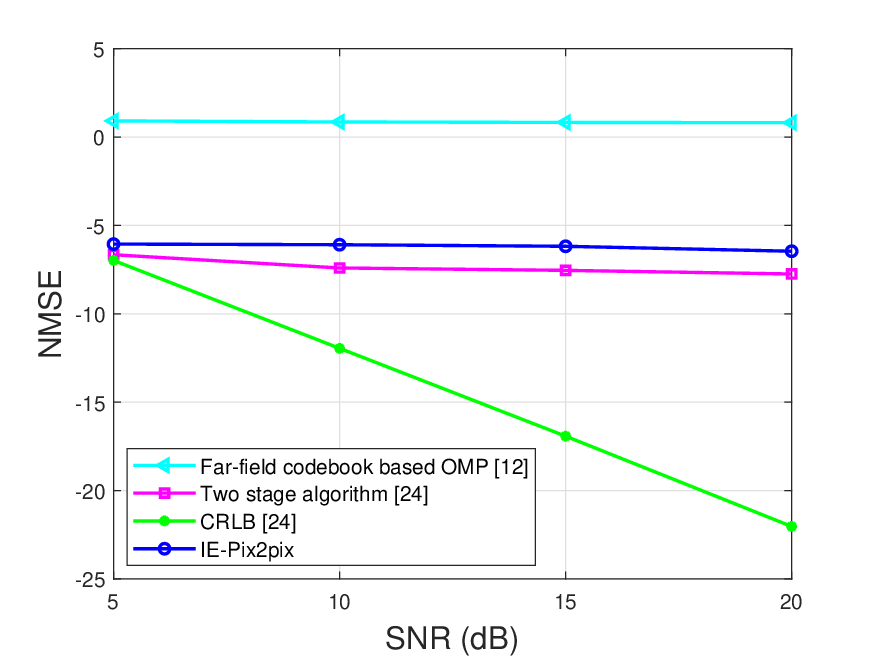}
		\subcaption{}
	\end{minipage}	
	
	\caption{NMSE performance versus SNR for different schemes.
		(a) $r=$ 40 m.	(b) $r=$ 60 m.}
\end{figure*}

As shown in Fig. 8 (b), the proposed IE-Pix2pix maintains good NMSE performance when the pilot length significantly decreases. In Fig 8 (b), the distance is 60 m. 
The IE-Pix2pix outperforms the two stage algorithm and the CRLB when the pilot length is smaller than 32. When the pilot length is decreased from 32 to 16, the NMSE performance of the two stage algorithm degrades dramatically. In the meanwhile, increasing the pilot length from 64 to 128 only provides marginal performance improvement for the two stage algorithm.
This result demonstrates that the enough measurement data diversity is provided for the algorithm as the pilot length increases to a certain extent.

Fig. 9 shows the NMSE performance comparison of different methods against different SNRs.
The size of the pilot matrix is $256\times 128$.
It can be observed from Fig. 9 (a) that the IE-Pix2pix outperforms the two stage algorithm in the range of all the SNRs when the distance $r=$ 40 m.
Furthermore, the IE-Pix2pix has better NMSE performance than the CRLB when SNR $\leqslant$ 7 dB.
With the increase in SNR, the NMSE performance of the IE-Pix2pix and the two stage algorithm improves very slightly. 
In addition, the IE-Pix2pix achieves much better performance than the far-field codebook based OMP scheme.
In Fig 9 (b), the distance is 60 m.
Although the NMSE performance of the proposed IE-Pix2pix improves very slightly, the IE-Pix2pix has much better performance than the far-field codebook based OMP approach. 

\section{Conclusion}

In this paper, we proposed a channel estimation method based on the IE-Pix2pix for XL-MIMO communication systems. 
In particular, the Pix2pix was developed to simultaneously estimate the LoS and NLoS path components of the XL-MIMO near-field channel. Moreover, we combined the adversarial loss introduced by the GAN architecture with a traditional loss function to ensure the right direction of training the generator. We also investigated the influence of the hyper-parameter of the loss function on the NMSE performance of the proposed method. To estimate the XL-MIMO near-field channels more accurately, the initially estimated channels instead of the received signals were fed into the designed Pix2pix. 
Numerical results showed that the proposed IE-Pix2pix based channel estimation method outperformed the two stage algorithm and the far-field codebook based OMP scheme in the adopted mixed LoS/NLoS XL-MIMO near-field channel model. 
Furthermore, the IE-Pix2pix surpassed the CRLB when the distance between the transmitter and the receiver was small and the pilot overhead was low.

\vspace{12pt}

\ifCLASSOPTIONcaptionsoff
  \newpage
\fi



%


\begin{thebibliography}{00}
 
	
\bibitem{b1} M. Cui, Z. Wu, Y. Lu, X. Wei, and L. Dai, ``Near-field MIMO communications for 6G: Fundamentals, challenges, potentials, and future directions,"  \textit{IEEE Commun. Mag.}, vol. 61, no. 1, pp. 40-46, Jan. 2023.
	
	
	
		
	
\bibitem{b2} P. P. Ray, N. Kumar, and M. Guizani, ``A vision on 6G-enabled NIB: Requirements, technologies, deployments, and prospects," \textit{IEEE Wireless Commun.}, vol. 28, no. 4, pp. 120-127, Aug. 2021.

	
\bibitem{b3} M. Giordani, M. Polese, M. Mezzavilla, S. Rangan, and M. Zorzi, ``Toward 6G networks: Use   cases and technologies," \textit{IEEE Commun. Mag.}, vol. 58, no. 3, pp. 55-61, Mar. 2020.








	
\bibitem{b4} K. T. Selvan and R. Janaswamy, ``Fraunhofer and Fresnel distances: Unified derivation for aperture antennas," \textit{IEEE Antennas and Propag. Mag.}, vol. 59, no. 4, pp. 12-15, Aug. 2017.
	
	
\bibitem{b5} C. Feng, H. Lu, Y. Zeng, T. Li, S. Jin, and R. Zhang, ``Near-field modelling and performance analysis for extremely large-scale IRS communications," \textit{IEEE Trans. Wireless Commun.}, to be published.
	
	
\bibitem{bb6} Z. Zhou, X. Gao, J. Fang, and Z. Chen, ``Spherical wave channel and analysis for large linear array in LoS conditions," in \textit{Proc. IEEE Globecom Workshops (GC Wkshps)}, Dec. 2015, pp. 1-6.








\bibitem{bb7} B. Friedlander, ``Localization of signals in the near-field of an antenna array," \textit{IEEE Trans. Signal Process.}, vol. 67, no. 15, pp. 3885-3893, Aug. 2019.


	
	
	
\bibitem{b6} H. Lu and Y. Zeng, ``Near-field modeling and performance analysis for multi-user extremely large-scale MIMO communication," \textit{IEEE Commun. Lett.}, vol. 26, no. 2, pp. 277-281, Feb. 2022.



	
	
	
	
\bibitem{b7} X. Yin, S. Wang, N. Zhang, and B. Ai, ``Scatterer localization using large-scale antenna arrays based on a spherical wave-front parametric model," \textit{IEEE Trans. Wireless Commun.}, vol. 16, no. 10, pp. 6543-6556, Oct. 2017.





\bibitem{bb8} Y. Chen, Y. Wang, and Z. Wang, ``Reconfigurable intelligent surface aided high-mobility millimeter wave communications with dynamic dual-structured sparsity," \textit{IEEE Trans. Wireless Commun.}, vol. 22, no. 7, pp. 4580-4599, Jul. 2023.



\bibitem{bb9} G. Zhou, C. Pan, H. Ren, P. Popovski, and A. L. Swindlehurst, ``Channel estimation for RIS-aided multiuser millimeter-wave systems," \textit{IEEE Trans. Signal Process.}, vol. 70, pp. 1478-1492, Mar. 2022.



\bibitem{bb10} J. Lee, G. -T. Gil, and Y. H. Lee, ``Channel estimation via orthogonal matching pursuit for hybrid MIMO systems in millimeter wave communications," \textit{IEEE Trans. Commun.}, vol. 64, no. 6, pp. 2370-2386, Jun. 2016.



\bibitem{b8} X. Wei, C. Hu, and L. Dai, ``Deep learning for beamspace channel estimation in millimeter-wave massive MIMO systems," \textit{IEEE Trans. Commun.}, vol. 69, no. 1, pp. 182-193, Jan. 2021.




\bibitem{b9} X. Gao, L. Dai, S. Zhou, A. M. Sayeed, and L. Hanzo, ``Wideband beamspace channel estimation for millimeter-wave MIMO systems relying on lens antenna arrays," \textit{IEEE Trans. Signal Process.}, vol. 67, no. 18, pp. 4809-4824, 15 Sep.15, 2019.




\bibitem{b10} C. Hu, L. Dai, T. Mir, Z. Gao, and J. Fang, ``Super-resolution channel estimation for mmWave massive MIMO with hybrid precoding," \textit{IEEE Trans. Veh. Technol.}, vol. 67, no. 9, pp. 8954-8958, Sep. 2018.



\bibitem{b11} Y. Han, S. Jin, C. -K. Wen, and X. Ma, ``Channel estimation for extremely large-scale massive MIMO systems," \textit{IEEE Wireless Commun. Lett.}, vol. 9, no. 5, pp. 633-637, May 2020.



\bibitem{b12} M. Cui and L. Dai, ``Channel estimation for extremely large-scale MIMO: Far-field or near-field?," \textit{IEEE Trans. Commun.}, vol. 70, no. 4, pp. 2663-2677, Apr. 2022.



\bibitem{b13} Z. Tang, Y. Chen, Y. Wang, T. Mao, Q. Wu, M. D. Renzo, and L. Hanzo, ``Near-field sparse channel estimation for extremely large-scale RIS-aided wireless communications," in \textit{Proc. IEEE Global Commun. Conf. (GLOBECOM)}, to be published.





\bibitem{b14} X. Zhang, Z. Wang, H. Zhang, and L. Yang, ``Near-field channel estimation for extremely large-scale array communications: A model-based deep learning approach," \textit{IEEE Commun. Lett.}, vol. 27, no. 4, pp. 1155-1159, Apr. 2023.

	
 
	
 

 
	

	
	

	
	
\bibitem{b15} A. Alkhateeb, O. El Ayach, G. Leus, and R. W. Heath, ``Channel estimation and hybrid precoding for millimeter wave cellular systems," \textit{IEEE J. Sel. Topics in Signal Process.}, vol. 8, no. 5, pp. 831-846, Oct. 2014.





	
\bibitem{bb16} Z. Gong, C. Li, F. Jiang, and M. Z. Win, ``Data-aided Doppler compensation for high-speed railway communications over mmWave bands," \textit{IEEE Trans. Wireless Commun.}, vol. 20, no. 1, pp. 520-534, Jan. 2021.
	

\bibitem{bb17} X. Wei and L. Dai, ``Channel estimation for extremely large-scale massive MIMO: Far-field, near-field, or hybrid-field?," \textit{IEEE Commun. Lett.}, vol. 26, no. 1, pp. 177-181, Jan. 2022.	


\bibitem{bb18} Z. Hu, C. Chen, Y. Jin, L. Zhou, and Q. Wei, "Hybrid-field channel estimation for extremely large-scale massive MIMO system," \textit{IEEE Commun. Lett.}, vol. 27, no. 1, pp. 303-307, Jan. 2023.





\bibitem{b16} Y. Lu and L. Dai, ``Near-field channel estimation in mixed LoS/NLoS environments for extremely large-scale MIMO systems," \textit{IEEE Trans. Commun.}, vol. 71, no. 6, pp. 3694-3707, Jun. 2023.

	

\bibitem{b18} I. Goodfellow \textit{et al.}, ``Generative adversarial nets," in \textit{Proc. NIPS}, Dec. 2014, pp. 2672-2680. 
	

\bibitem{b19} W. Chen and J. Hays, ``SketchyGAN: Towards diverse and realistic sketch to image synthesis," in \textit{Proc. IEEE Conf. Comput. Vis. Pattern Recognit. (CVPR)}, Jun. 2018, pp. 9416-9425.



\bibitem{b20} H. Tang, H. Liu, and N. Sebe, ``Unified generative adversarial networks for controllable image-to-image translation," \textit{IEEE Trans. Image Process.}, vol. 29, pp. 8916-8929, 2020. 

	
 


\bibitem{b21} A. S. Doshi, M. Gupta, and J. G. Andrews, ``Over-the-air design of GAN training for mmWave MIMO channel estimation," \textit{IEEE J. Sel. Areas
Inf. Theory}, vol. 3, no. 3, pp. 557-573, Sep. 2022.

\bibitem{bb22} E. Balevi and J. G. Andrews, ``Wideband channel estimation with a generative adversarial network," \textit{IEEE Trans.   Wireless Commun.}, vol. 20, no. 5, pp. 3049-3060, May 2021.	



\bibitem{b22} M. Ye, H. Zhang, and J. -B. Wang, ``Channel estimation for intelligent reflecting surface aided wireless communications using conditional GAN," \textit{IEEE Commun. Lett.}, vol. 26, no. 10, pp. 2340-2344, Oct. 2022.


\bibitem{b23} Y. Dong, H. Wang, and Y. D. Yao, ``Channel estimation for one-bit multiuser massive MIMO using conditional GAN," \textit{IEEE Commun. Lett.}, vol. 25, no. 3, pp. 854-858, Mar. 2021.    


\bibitem{b24} T. Hu, Y. Huang, Q. Zhu, and Q. Wu, ``Channel estimation enhancement with generative adversarial networks,"\textit{IEEE Trans. Cogn. Commun. Net.}, vol. 7, no. 1, pp. 145-156, Mar. 2021.


 
	
	
	
\bibitem{b25} P. Isola, J. Zhu, T. Zhou, and A. A. Efros, ``Image-to-image translation with conditional adversarial networks," in \textit{Proc. IEEE Conf. Comput. Vis. Pattern Recognit. (CVPR)}, Honolulu, HI, USA, Jul. 2017, pp. 5967-5976.



	
 

	
\bibitem{b26} J. Sherman, ``Properties of focused apertures in the Fresnel region," \textit{IRE Trans. Antennas and Propag.}, vol. 10, no. 4, pp. 399-408, Jul. 1962.



	
	



	
	 
	


 







	
	
 
	
	
	

	


\bibitem{b27} H. Ye, L. Liang, G. Y. Li, and B. Juang, ``Deep learning-based end-to-end wireless communication systems with conditional GANs as unknown channels," \textit{IEEE Trans. Wireless Commun.}, vol. 19, no. 5, pp. 3133-3143, May 2020.



	

	
	

	
	
	



	
\bibitem{b28} P. Dong, H. Zhang, G. Y. Li, I. S. Gaspar, and N. NaderiAlizadeh, ``Deep CNN-based channel estimation for mmWave massive MIMO systems," \textit{IEEE J. Sel. Topics Sig. Process.}, vol. 13, no. 5, pp. 989-1000, Sep. 2019.
	
	
	
	

	
	
	
	
	
	
	
	
	
	
	
	
	
\end{thebibliography}

\vspace{-3cm}
\begin{IEEEbiography}
	[{\includegraphics[width=1.05in,height=1.25in,clip,keepaspectratio]{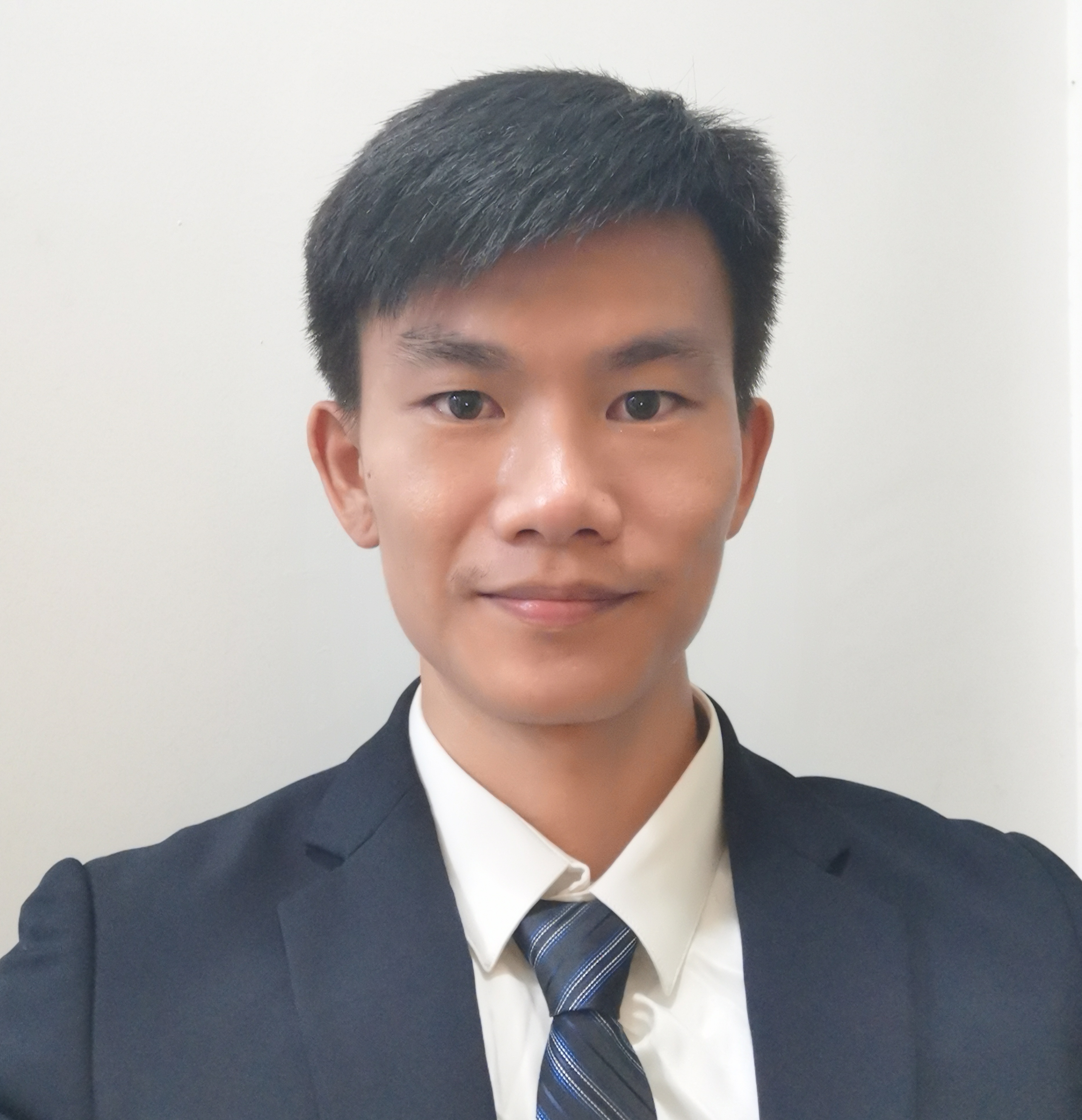}}] 
	{Ming Ye} (Graduate Student Member, IEEE)
	received the B.S. degree in communication engineering
	from Jilin University, Changchun, China, in 2016,
	and the M.S. degree in information and communication engineering from
	Hainan University, Haikou, China, in 2019.
	He is currently
	pursuing the Ph.D. degree at the National Mobile
	Communications Research Laboratory, Southeast
	University, Nanjing, China. His
	research interests include intelligent reflection surface, AI for wireless, and near field communications.
\end{IEEEbiography}
\vspace{-2cm}
\begin{IEEEbiography}
[{\includegraphics[width=1.05in,height=1.25in,clip,keepaspectratio]{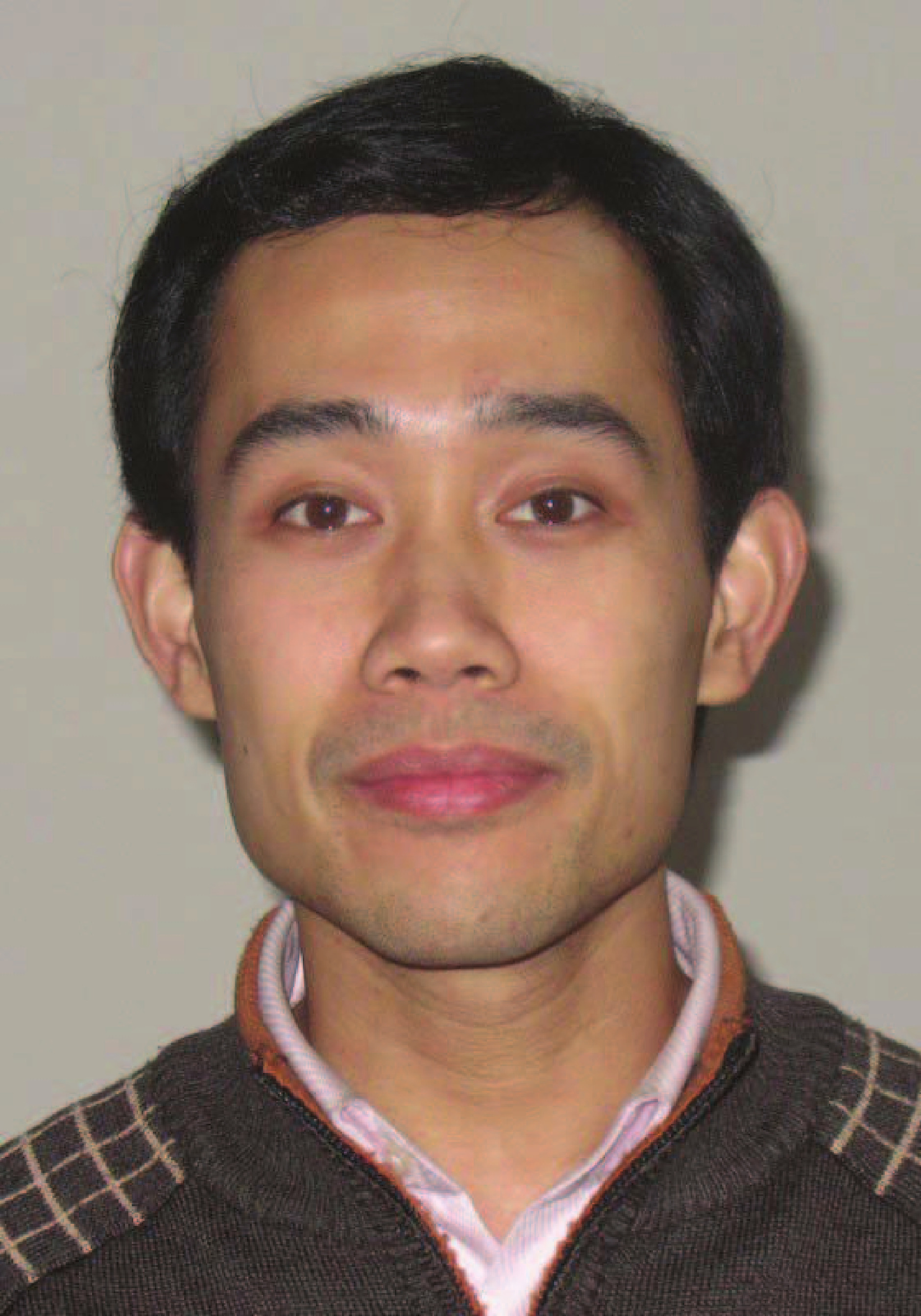}}] 
{Xiao Liang} (Member, IEEE) received his B.Sc., M.S. and Ph.D. degrees in Communication and Information Engineering in 2000, 2005 and 2013, respectively, all from Southeast University.
He is currently a lecturer with National Mobile Communications Research Lab., (NCRL) Southeast University, 
and also with the Purple Mountain Laboratories.  His research interests lie in future-generation wireless communications and networks, including signal processing, machine learning, wireless networking,  and optical wireless communications.
\end{IEEEbiography}
\vspace{-3cm}
\begin{IEEEbiography}
	[{\includegraphics[width=1in,height=1.25in,clip,keepaspectratio]{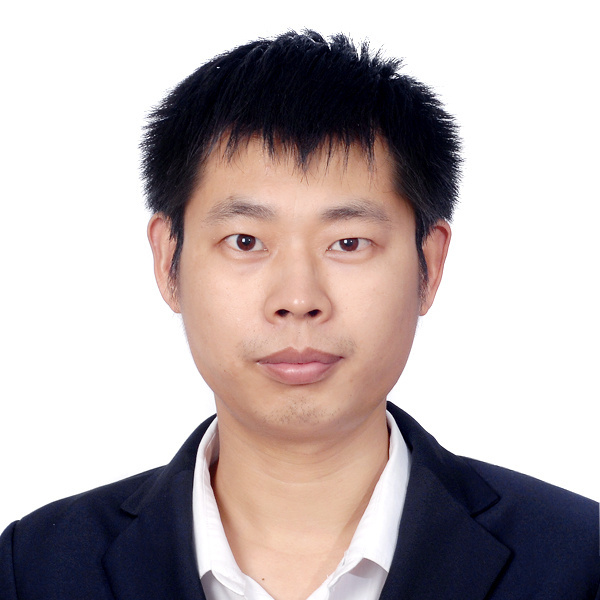}}] 
	{Cunhua Pan} (Senior Member, IEEE) is a full professor in Southeast University.  His research interests mainly include  reconfigurable intelligent surfaces (RIS),  AI for Wireless, and near field communications and sensing. He has published over 170 IEEE journal papers. His papers got over 14,000 Google Scholar citations with H-index of 60. He is  Clarivate Highly Cited researcher. He is/was an Editor of IEEE Transaction on Communications, IEEE Transactions on Vehicular Technology, IEEE Wireless Communication Letters, IEEE Communications Letters and IEEE ACCESS. He serves as the guest editor for IEEE Journal on Selected Areas in Communications on the special issue on xURLLC in 6G: Next Generation Ultra-Reliable and Low-Latency Communications. He also serves as a leading guest editor of IEEE Journal of Selected Topics in Signal Processing (JSTSP)  Special Issue on Advanced Signal Processing for Reconfigurable Intelligent Surface-aided 6G Networks, leading guest editor of IEEE Vehicular Technology Magazine on the special issue on Backscatter and Reconfigurable Intelligent Surface Empowered Wireless Communications in 6G, leading guest editor of IEEE Open Journal of Vehicular Technology on the special issue of Reconfigurable Intelligent Surface Empowered Wireless Communications in 6G and Beyond, and leading guest editor of IEEE IEEE Transactions on Green Communications and Networking Special Issue on Design of Green Near-Field Wireless Communication Networks.   He received the  IEEE ComSoc Leonard G. Abraham Prize in 2022 and IEEE ComSoc Asia-Pacific Outstanding Young Researcher Award, 2022. 
\end{IEEEbiography}
\vspace{-1cm}

\begin{IEEEbiography}
	[{\includegraphics[width=1in,height=1.25in,clip,keepaspectratio]{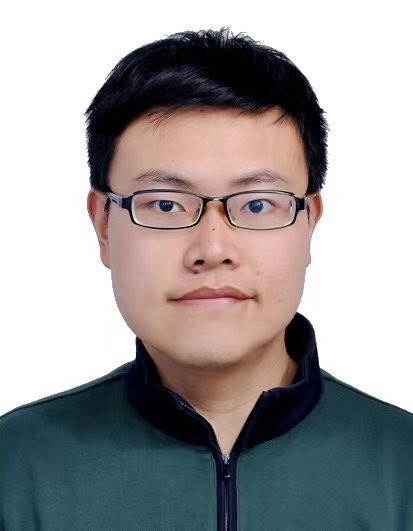}}] 
	{Yinfei Xu} (Member, IEEE) received the B.E. and Ph.D. degrees in information engineering from Southeast University, Nanjing, China, in 2008 and 2016, respectively. From July 2014 to January 2015, he was a Visiting Student with the Department of Electrical and Computer Engineering, McMaster University, Hamilton, ON, Canada. From March 2016 to July 2017, he was also a Research Assistant and a Post-Doctoral Fellow with the Institute of Network Coding, The Chinese University of Hong Kong, Hong Kong. Since August 2017, he has been with the School of Information Science and Engineering, Southeast University, where he is currently an Associate Professor. His research interests include information theory, wireless communications, and machine learning.
\end{IEEEbiography}
\vspace{2cm}

\begin{IEEEbiography}
	[{\includegraphics[width=1in,height=1.25in,clip,keepaspectratio]{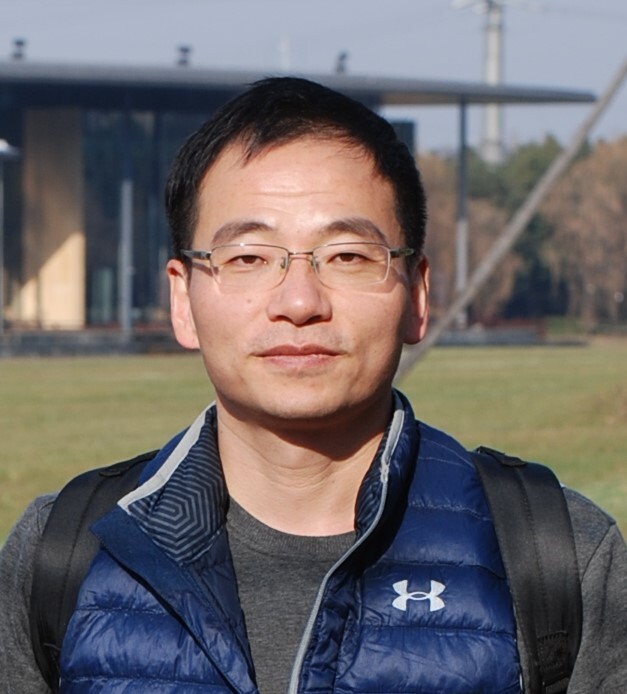}}] 
	{Ming Jiang} Ming Jiang (Member, IEEE) received the B.Sc.,
	M.S., and Ph.D. degrees in communication and
	information engineering from Southeast University,
	Nanjing, China, in 1998, 2003, and 2007, respectively.
	He is currently an Associate Professor with
	the National Mobile Communications Research Laboratory,
	Southeast University. His research interests
	include coding and modulation technology.
\end{IEEEbiography}

\vspace{2cm}

\begin{IEEEbiography}
	[{\includegraphics[width=1in,height=1.25in,clip,keepaspectratio]{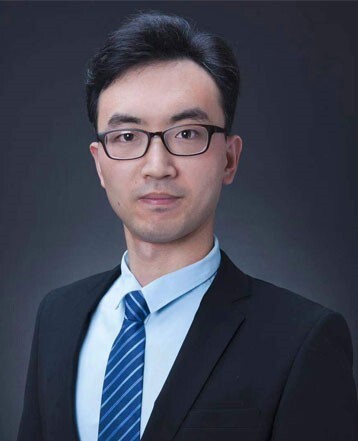}}] 
	{Chunguo Li} (SM'16) received the bachelor's degree in wireless communications from Shandong University in 2005, and Ph.D. degree in wireless communications from Southeast University in 2010. In July 2010, he joined the Faculty of Southeast University, Nanjing China, where he was Associate Professor between 2012 and 2016, and Full Professor since 2017 to present. From June 2012 to June 2013, he was the Postdoctor with Concordia University, Montreal, Canada. From July 2013 to August 2014, he was with the DSL laboratory of Stanford University as Visiting Associate Professor. From August 2017 to July 2019, he was the adjunct professor of Xizang Minzu University under the supporting Tibet program organized by China National Human Resources Ministry. 
	He is the Fellow of IET, Fellow of China Institute of Communications (CIC), Chair of IEEE Computational Intelligence Society Nanjing Chapter, and Chair of Advisory Committee for Instruments industry in Jiangsu province. He has served as editor for a couple of international journals and as session chair for many international conferences. His research interests are in 6G cell-free distributed MIMO wireless communications, information theories, and AI based audio signal processing. 
\end{IEEEbiography}

\end{document}